\documentclass[10pt]{article}
\usepackage[inner=3cm,outer=2cm]{geometry}
\usepackage{graphicx}
\usepackage{amssymb}
\usepackage{mathtools} 

\newenvironment{dfn}{\refstepcounter{thm} \bigskip \par \noindent
    {\bf Definition \thethm}\ }{\bigskip \par}
\newenvironment{exa}{\refstepcounter{thm} \bigskip \par \noindent
        {\bf Example \thethm}\ }{\bigskip \par}

 \DeclareMathOperator*{\vol}{Volume}

\begin{document}
\title{Distribution-Free Bayesian multivariate predictive inference}

\author{Daniel Yekutieli}

\baselineskip=15pt

\maketitle

\abstract{
We introduce a comprehensive Bayesian multivariate predictive inference framework.
The basis for our framework is a hierarchical Bayesian model, 
that is a mixture of finite Polya trees corresponding to multiple dyadic partitions of the unit cube.
Given a sample of observations from an unknown multivariate distribution,
the posterior predictive distribution is used to model and generate future observations from the unknown distribution.
We illustrate the implementation of our methodology and study its performance on simulated examples.
We introduce an algorithm for constructing conformal prediction sets, 
that provide finite sample probability assurances for future observations, with our Bayesian model.
}

\section{Introduction}
The predictive distribution is an important feature of Bayesian analyses that is typically used as a diagnostic tool for assessing model fit.
In this work we apply a hierarchical Bayesian model, 
that is a mixture of finite Polya trees corresponding to multiple dyadic partitions of  $[0,1]^P$,
to a sample of iid observations from an unknown distribution $\tilde{\pi}$.
The resulting posterior predictive distribution is used to approximate the distribution of future observations from $\tilde{\pi}$.

The use of finite Polya trees to define random distributions on dyadic partitions of the unit interval was introduced in Ferguson (1974).
The theoretical properties of Polya tree models for density estimation has been widely studied.
Use of  Polya trees, corresponding to multiple dyadic partitions, for distribution-free estimation of multivariate densities was suggested in Wong and Ma (2010) and Ma (2017).
The novel features of this work is the introduction of a hierarchical Bayesian model that assigns distributions to sets of dyadic partitions of $[0,1]^P$;
the introduction and implementation of a comprehensive Bayesian predictive inference framework;
the adaptation of the Vovk (2005) method for constructing Conformal prediction sets to Bayesian analyses.

\medskip
In Section 2 we present our inferential framework.
In Sections 3, 4, 5, we illustrate the use of our methodology and study its performance on simulated examples.
In Section 3 we study how well our model approximates and estimates one-dimensional and two-dimensional densities.
In Section 4 we apply our method for quantile regression for a two-dimensional density,
and introduce and implement an algorithm for constructing Conformal prediction sets with our Bayesian model.
In Section 5 we attempt to scale up our methodology to provide predictive inference for a sample of $8$ continuous variables and one categorical variable.
In Section 6 we propose modifications for our framework for the case that continuous variables need to be partitioned into unequally probable subintervals or 
for modeling  categorical variables.
For equal volume segmentation we use the same value $a_0$ for the the two Beta prior hyper parameters.
 For the unequal volume setting we suggest changing the hyper parameters odds from $0.5$ to 
 the corresponding data adaptive volume odds. 
 In Section 7 we discuss our work and outline changes needed to make it applicable in high dimensional problems.

\section{Hierarchical Bayesian modelling for predictive inference}

\subsection{Dyadic segmentations of the $P$-dimensional unit cube}

$L$ level dyadic segmentations are constructed by partitioning the unit cube, ${\cal I}_0 =  [0,1]^P$, $L$ times  in half according to its $P$ dimensions.
Each segmentation,  $\vec{\cal I} = \{ {\cal I}_{l,j} \}_{l = 1 \cdots L}^{j = 1 \cdots 2^{l}}$, 
consists of $2 \cdot (2^L - 1)$ subintervals specified by a segmentation ordering vector,  $\vec{d} = \{ d_{l,j} \}_{l = 1 \cdots L}^{j = 1 \cdots 2^{l-1}}$ with $d_{l,j} \in \{ 1, \cdots, P \}$.
The segmentation begins by partitioning ${\cal I}_0$ according to Dimension $d_{1,1}$ into two halves,
${\cal I}_{1,1} (\vec{d})$ that consists of the small values of coordinate $d_{1,1}$ and ${\cal I}_{1,2}  (\vec{d})$ that consists of the large values of coordinate $d_{1,1}$. 
And then, for $l = 2 \cdots L$ and $j = 1 \cdots 2^{l-1}$,  
${\cal I}_{l-1,j}  (\vec{d})$ is partitioned in halve according to  coordinate $d_{l,j}$ into  ${\cal I}_{l, 2 \cdot j - 1}  (\vec{d})$ consisting of  small values of coordinate $d_{l, j}$
and  into ${\cal I}_{l, 2 \cdot j}  (\vec{d})$ consisting of large values of coordinate $d_{l, j}$.

In the segmentations we consider in the examples in this report, 
all the subintervals at the same level are partitioned according to the same dimension (i.e. $d_{l,j}$ is the same $\forall j$).
In which case,  $\vec{d}$ is given by $(d_1 \cdots d_L)$.

\medskip
In Figure 1 we display four $L=4$ level dyadic segmentations of $[0,1]^2$ into $16 = 2^4$ subintervals.
Setting the X axis to be  Dimension 1 and the Y axis to be  Dimension 2,
the segmentation with all $d_{l,j} = 1$ is denoted  by  $(X, X, X, X)$;
the segmentation with all $d_{l,j} = 2$ is denoted  by  $(Y, Y, Y, Y)$;
while the segmentation with $d_{1,j} = 1$, $d_{2, j} = 1$, $d_{3,j} = 2$ and $d_{4,j} = 2$ is denoted $(X, X, Y, Y)$.
The numbers in each subinterval mark the positions of the 
level $L = 4$ subintervals, ${\cal I}_{4,1} (\vec{d}) \cdots {\cal I}_{4, 16} (\vec{d})$, for each segmentation.

\begin{figure} [p] \label{figure1}
\center
\includegraphics[width=1\textwidth, height=.6\textwidth]{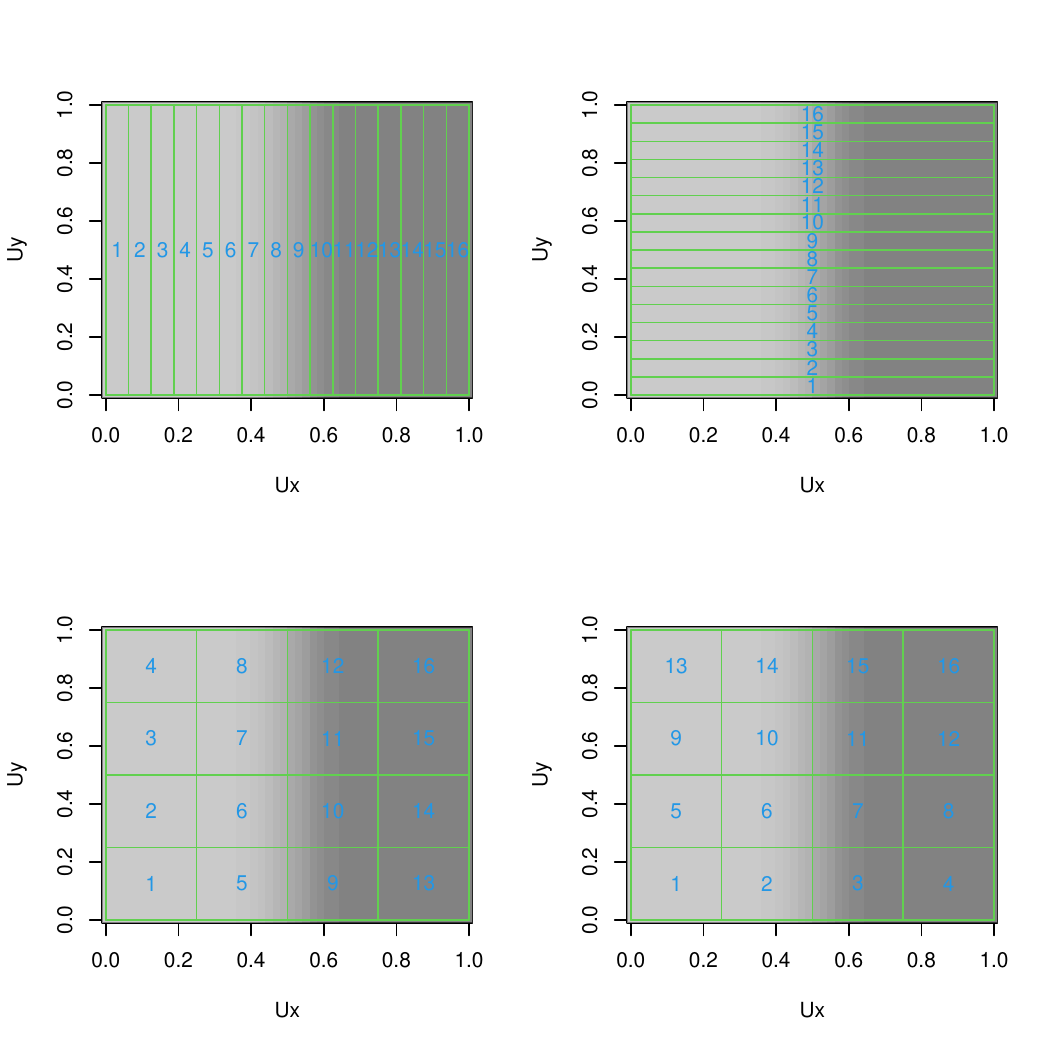}
\caption{$4$ level dyadic segmentations of $[0,1]^2$.
The top left plot displays segmentation $(X, X, X, X)$;
the top right plot displays segmentation $(Y, Y, Y, Y)$;
the bottom left plot displays segmentation $(X, X, Y, Y)$;
the bottom right plot displays segmentation $(Y, Y, X, X)$.
The blue numbers in each plot mark the positions of the 
level $4$ subintervals, ${\cal I}_{4,1} \cdots {\cal I}_{4, 16}$, for each segmentation.
The gray shading in all the plots displays density $\tilde{\pi} ( u_x,u_y) = 2 \cdot logit^{-1} (20 \cdot (u_x-0.5))$.
}
\end{figure}

\begin{figure}[p] 
\centering
\includegraphics[width=1\textwidth,height=.3\textwidth]{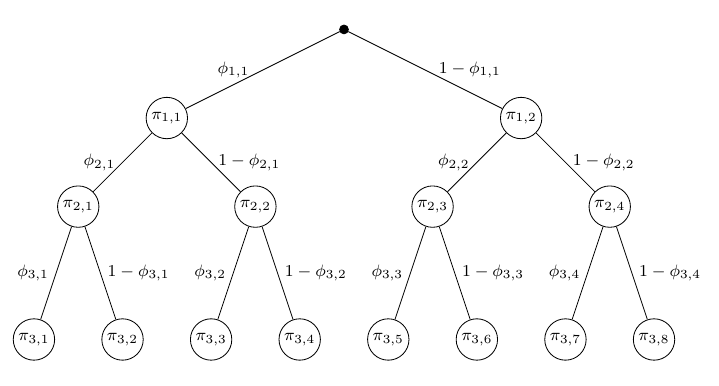}
\caption{$3$ level  hierarchical Beta model. 
$\phi_{1,1},  \phi_{2,1},  \cdots,  \phi_{3,4}$ are independent Beta random variables.
For $l = 1\cdots 3$ and $j = 1\cdots 2^l$,  the probability $\pi_{l, j}$ is the product of the Beta random variables 
on the edges connecting it to the root node (solid black circle). 
The probabilities at level $3$ specify the step function PDF for a dyadic partition of $[0,1]^P$.
}
\label{hBeta_scehamtic}
\end{figure}

\bigskip
\subsection{The hierarchical Beta model}
The $L$ level hierarchical Beta  (hBeta) model is a finite Polya tree model that generates random densities that are step 
functions on  $\{ {\cal I}_{L,1}, \cdots, {\cal I}_{L, 2^L} \}$ for a given $L$ level segmentation $\vec{d}$.
The parameters of the hBeta  model are Beta distribution hyper-parameters, $\alpha_{l, j}$ and  $\beta_{l, j}$ for $l = 1 \cdots L$ and $j = 1 \cdots 2^{l-1}$.
 The  hBeta model generates the following components.

\medskip
\noindent {\bf a. Independent Beta random variables.}
 $ \phi_{l, j} \sim Beta ( \alpha_{l, j}, \beta_{l, j})$, for $l = 1 \cdots L$ and $j = 1 \cdots 2^{l-1}$.
 The Beta random variables specify the conditional subinterval probabilities.
$\Pr({\cal I}_{1, 1} | {\cal I}_0  ) = \phi_{1, 1}$ and $\Pr({\cal I}_{1,2}  | {\cal I}_0 )  = 1 - \phi_{1, 1}$.
 For  $l = 2 \cdots L$ and $j = 1 \cdots 2^{l-1}$, $\Pr({\cal I}_{l, 2 \cdot j - 1} | {\cal I}_{l-1, j}) = \phi_{l, j}$ and $\Pr({\cal I}_{l, 2 \cdot j } | {\cal I}_{l-1, j}) = 1 - \phi_{l, j}$.

\medskip
\noindent {\bf b. Subinterval probabilities.}  The subinterval probabilities, $\Pr({\cal I}_{l,j})  = \pi_{l, j}$,
are products of the conditional subinterval probabilities.
$\pi_{1, 1} =  \phi_{1, 1}$ and $\pi_{1, 2} = 1 - \phi_{1, 1}$.
For $l = 2 \cdots L$ and $j = 1 \cdots 2^{l-1}$,
$\pi_{l,2 \cdot  j -1} = \phi_{l,  j} \cdot \pi_{l-1,j}$ and $\pi_{l,2 \cdot  j} = (1- \phi_{l,  j}) \cdot \pi_{l-1,j}$.

\medskip
\noindent {\bf c. Step function PDF.}
The components of $\vec{\pi}_L = (\pi_{L, 1} \cdots \pi_{L, 2^L})$ specify the step function PDF, 
\begin{equation} \label{def-step}
f ( u  |  \vec{d}, \vec{\pi}_L )  =  \pi_{L,1} \cdot  \frac{ I_{L,1} (u ; \vec{d}) }{(1/2)^L} +  \cdots +
\pi_{L, 2^L} \cdot \frac{ I_{L, 2^L} (u  ; \vec{d}) }{ (1/2)^L},
 \end{equation}
for $u \in [0,1]^P$ and with $I_{l, j} (u ; \vec{d})$ denoting the indicator function corresponding to subinterval ${\cal I}_{l,j} (\vec{d})$.

\medskip \noindent
In Figure \ref{hBeta_scehamtic} we provide a schematic for the hierarchical Beta model with $L=3$ levels.

\bigskip
\subsection{The hierarchical Bayesian framework}
Let  $u_1 \cdots  u_{m}$ denote the sample of  iid observations from $\tilde{\pi}$.
On observing data points $u_1 \cdots u_{m}$, our goal is to provide inference regarding an unobserved data point $u_{m+1}$,
that is assumed to be also independently sampled from  $\tilde{\pi}$.
To drive our inferential  framework,
we elicit the following generative model for $u_1 \cdots u_{m+1}$.

\begin{dfn} {\bf Generative model for data} \label{gen:model}
 \begin{enumerate}
\item Sample $\vec{d}$ with probability $1 / | {\cal D} |$  from a given set of level $L$ segmentations ${\cal D}$.

\item Generate $ f ( u  |  \vec{d}, \vec{\pi}_L  )$ from the hBeta model with $\phi_{l,j} \sim Beta(1,1)$, for $l = 1 \cdots L$ and  $j = 1 \cdots 2^{l-1}$.

\item For $k = 1 \cdots m+1$, generate  $u_k  \sim  f ( u | \vec{d}, \vec{\pi}_L )$.
 \end{enumerate}
\end{dfn}

\begin{exa} \label{exa111}
In Figure 3 we display the distribution of the CDF of  $f ( u | \vec{d}, \vec{\pi}_L )$ in Model  \ref{gen:model},
for the $10$ level segmentation of $[0,1]$, with  subintervals ${\cal I}_{10, j} = [ \frac{j-1}{1024} , \frac{j}{1024} ]$ for $j = 1 \cdots 1024$, for $a_0 = 0.1, 1, 10$.  
In Model  \ref{gen:model}, $\phi_{l,j}$ are iid with $E \phi_{l,j} = a_0 / (a_0 + a_0) = 1/2$,
therefore $E \pi_{L,j} = (1/2)^L$, and thus the expectation of $f ( u | \vec{d}, \vec{\pi}_L )$ is the $U[0,1]$ density.

The plots reveal that increasing $a_0$ decreases the dispersion of $f ( u | \vec{d}, \vec{\pi}_L )$.
For $a_0 = 0.1$, the probabilities in $\vec{\pi}_{10}$ are concentrated --  very unevenly -- in a few very small clusters of contiguous subintervals.
For $a_0 = 10$,  the distribution of $\vec{\pi}_{10}$  is spread out in large clusters of similarly probable contiguous subintervals.
While for $a_0 = 1$,  the distribution of $\vec{\pi}_{10}$  consists of large clusters of low probability contiguous subintervals and a few small clusters of contiguous high probability subintervals.
\end{exa}

\begin{figure}[h]  \label{plt111}
\centering
\includegraphics[width=1\textwidth,height=.4\textwidth]{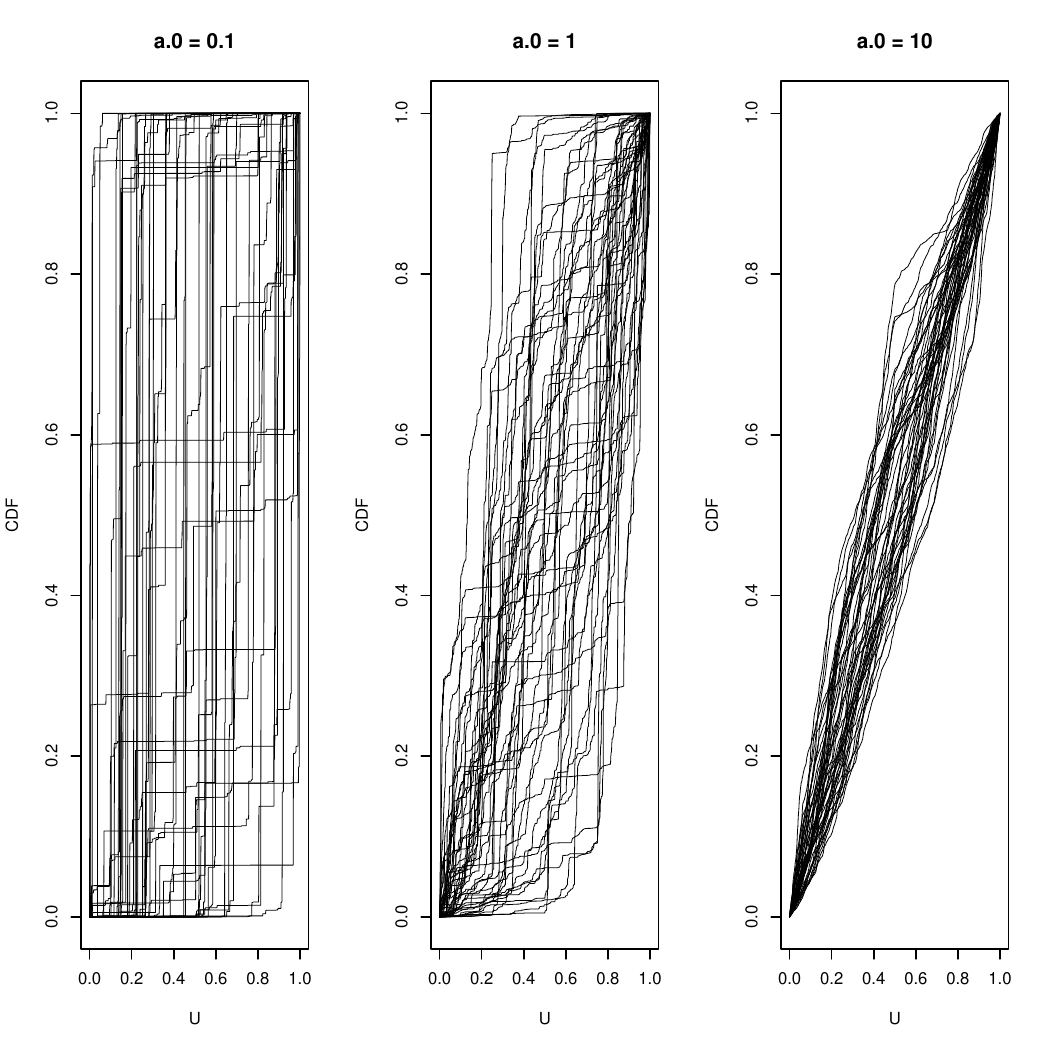}
\caption{
Generative model distribution of the step-function density.
The plots display the CDF of  step function $f ( u | \vec{d}, \vec{\pi}_L )$ in Model \ref{gen:model},
for the $10$ level segmentation of $[0,1]$ with $a_0 = 0.1$ (left plot), $a_0 = 1$ (middle plot), $a_0 = 10$ (right plot).
Each plot consists of curves that display the CDF of $f ( u | \vec{d}, \vec{\pi}_{10} )$ for $50$ realizations of  $\vec{\pi}_{10}$.
The X coordinates for each curve are the endpoints of the subintervals of $[0,1]$: $0, \frac{1}{1024}, \ldots,\frac{1023}{1024},  1$.
The Y coordinates for each curve are the cumulative sums of $\vec{\pi}_{10}$: $0, \pi_{10,1}, \pi_{10,1} + \pi_{10,2}, \ldots,  1$.
}
\end{figure}

\subsubsection{The posterior data model}

In this subsection we derive the conditional distribution  given $\vec{u} = (u_1 \cdots u_m)$, of $\vec{d}$ and the vector of Beta parameters,
$\vec{\phi} = \{ \phi_{l,j} \}_{l = 1 \cdots L}^{j = 1 \cdots 2^{l-1}}$, for Model \ref{gen:model}.

\medskip
For $\vec{d} \in {\cal D}$, $k = 1 \cdots m$, $l = 1 \cdots L$,  $j = 1 \cdots 2^{l}$,
let $N^k_{l, j} = I_{l,j} (u_k ; \vec{d})$ denote the indicator variable for the event $u_k \in {\cal I}_{l,j} (\vec{d})$.
The number of observations in subinterval ${\cal I}_{l,j} (\vec{d})$ is $N_{l, j} (\vec{u} ; \vec{d})  = \sum_{k = 1}^{m} N^k_{l, j}$.
Conditional on $\vec{\phi}$,
$N_{1,1} \sim Binomial (m, \phi_{1,1})$ 
and $N_{1,2} = m - N_{1,1}$.
For $l = 2 \cdots L$ and $j = 1 \cdots 2^{l-1}$, conditioning on $\vec{\phi}$ and on $N_{l-1, j}$,
$N_{l,2 \cdot j - 1} \sim Binomial( N_{l-1,j} , \phi_{l, j})$ and $N_{l, 2 \cdot j  } = N_{l-1,j} - N_{l, 2 \cdot j - 1}$.
Recalling that $\phi_{l,j} \sim Beta (a_0, a_0)$ and denoting 
$\vec{N}_{\ast, \ast} = \{  N_{l,j}  \}^{j =1 \cdots 2^l}_{ l = 1 \cdots L}$,
this implies that   $\phi_{l, j} | \vec{N}_{\ast, \ast}, \vec{d}  \sim Beta( a_0 + N_{l, 2 \cdot j - 1}, a_0 + N_{l, 2 \cdot j})$.

\medskip
Ferguson (1974) has already noted the conjugacy of the conditional distribution of $\vec{\phi}$ given $\vec{d}$ and $\vec{u}$ for the hBeta model.
In our inferential framework, the segmentation is random and we consider the conditional distribution of $\vec{\phi}$ and $\vec{d}$ given $\vec{u}$.
Denoting
$ \vec{N}_{L, \ast} = \{  N_{L,j}  \}_{j = 1 \cdots 2^l}$ and $ \vec{N}^{\ast}_{L, \ast} = \{  N^k_{L,j}  \}^{j = 1 \cdots 2^l}_{k = 1 \cdots m}$,
we begin by deriving  the conditional distribution of $\vec{d}$.
\begin{eqnarray}  \label{post-dist}
\lefteqn{ \Pr ( \vec{d} | \vec{u} )  =  \frac{ f (  \vec{d} , \vec{u})}
{ f ( \vec{u} )} =  \frac{ f (  \vec{d} , \vec{u})}
{ \sum_{\vec{d'} \in {\cal D}} f ( \vec{d'} , \vec{u} )} } \nonumber \\
& =  &   \frac{ f (   \vec{N}^{\ast}_{L,\ast } ,  \vec{N}_{L,\ast}, \vec{d}, \vec{u})}
{\sum_{\vec{d'} \in {\cal D}} f (   \vec{N}^{\ast}_{L,\ast } ,  \vec{N}_{L,\ast}, \vec{d'} , \vec{u})}  \nonumber \\
& =  &   \frac{ f ( \vec{u} |  \vec{N}^{\ast}_{L,\ast } ,  \vec{N}_{L,\ast}, \vec{d}  )
\cdot \Pr (   \vec{N}^{\ast}_{L,\ast }  |   \vec{N}_{L,\ast}, \vec{d}  )
\cdot \Pr (   \vec{N}_{L,\ast} |  \vec{d}  ) \cdot \Pr ( \vec{d}  )}
{\sum_{\vec{d'} \in {\cal D}} f ( \vec{u} |  \vec{N}^{\ast}_{L,\ast } ,  \vec{N}_{L,\ast}, \vec{d'}  )
\cdot \Pr (   \vec{N}^{\ast}_{L,\ast }  |   \vec{N}_{L,\ast}, \vec{d'}  )
\cdot \Pr (   \vec{N}_{L,\ast} |  \vec{d'}  ) \cdot \Pr ( \vec{d'}  )} \nonumber \\
& =  &   \frac{ \Pr (   \vec{N}^{\ast}_{L,\ast }  |   \vec{N}_{L,\ast}, \vec{d}  )
\cdot \Pr (   \vec{N}_{L,\ast} |  \vec{d}  ) }
{\sum_{\vec{d'} \in {\cal D}}
\cdot \Pr (   \vec{N}^{\ast}_{L,\ast }  |   \vec{N}_{L,\ast}, \vec{d'}  )
\cdot \Pr (   \vec{N}_{L,\ast} |  \vec{d'}  ) }  \label{eq671}
\end{eqnarray}
where the equality in (\ref{eq671}) is because  $ \Pr ( \vec{d'}  ) \equiv 1 / | {\cal  D}|$ and
$f ( \vec{u} |  \vec{N}^{\ast}_{L,\ast } ,  \vec{N}_{L,\ast}, \vec{d'}) = ( 2^L)^m$.
The latter holds because  given $\vec{d'}$ and $ \vec{N}^{\ast}_{L,\ast }$,  the components of $ \vec{u}$ are independently and Uniformly distributed within their respective level $L$ subintervals,
and each level $L$ subintervals for each  $\vec{d'}$ has volume $(1/2)^L$.
There are 
$ { m \choose N_{L,1}, \cdots, N_{L, 2^L}} = m!  /  (\Pi_{j = 1}^{2^L} N_{L,j}!)$ different indicator vector instances $ \vec{N}^{\ast}_{L,\ast }$ that yield the same counts vector $\vec{N}_{L, \ast}$.
As  the components of $ \vec{u}$ are exchangeable in Model \ref{gen:model} then each instance of $ \vec{N}^{\ast}_{L,\ast }$ is equally probable, threfore
\begin{equation} \label{ inv-mult-coef}
\Pr (   \vec{N}^{\ast}_{L,\ast }  |   \vec{N}_{L,\ast}, \vec{d'}  ) =  \frac{ \Pi_{j = 1}^{2^L} N_{L,j}  ! }{ m!}.
 \end{equation}
Note that for each segmentation $\vec{d}$, $N_{1,1} \sim Beta-Binomial(m, a_0, a_0)$, $N_{1,2} = m - N_{1,1}$, 
and for $l = 2 \cdots L$ and $j = 1 \cdots 2^{l-1}$, $N_{l, 2 \cdot j - 1} | N_{l - 1, j} \sim Beta-Binomial ( N_{l-1, j}, a_0, a_0)$
and $N_{l, 2 \cdot j }  =  N_{l - 1, j}  - N_{l, 2 \cdot j - 1}$.
Denoting $ \vec{N}_{l, \ast} =  (N_{l,1}, \cdots, N_{l, 2^l})$, 
recursively invoking these relations we may express,
\begin{eqnarray}
\lefteqn{\Pr( \vec{N}_{L, \ast}   | \vec{d})  = \Pr( \vec{N}_{1, \ast}, \cdots, \vec{N}_{L,\ast}  | \vec{d}) }  \nonumber \\
& =  & \Pr( \vec{N}_{L,\ast}   | \vec{N}_{1,\ast}, \cdots, \vec{N}_{L-1,\ast},  \vec{d} ) \cdot
 \Pr( \vec{N}_{1,\ast}, \cdots, \vec{N}_{L-1,\ast}  | \vec{d} )  \nonumber \\
& = & 
 \Pr( \vec{N}_{L,\ast}  | \vec{N}_{L-1,\ast},  \vec{d} ) \cdot \Pr( \vec{N}_{L-1,\ast}   | \vec{N}_{1,\ast}, \cdots, \vec{N}_{L-2,\ast},  \vec{d} ) \cdot
\Pr( \vec{N}_{1,\ast}, \cdots, \vec{N}_{L-2,\ast}  | \vec{d} ) \nonumber \\
& = & 
 \Pr( \vec{N}_{L,\ast}  | \vec{N}_{L-1,\ast},  \vec{d} ) \cdot \Pr( \vec{N}_{L-1,\ast}   |  \vec{N}_{L-2,\ast},  \vec{d} ) \cdot \; \cdots \; \cdot
\Pr( \vec{N}_{2,\ast} | \vec{N}_{1,\ast},  \vec{d} )  \cdot \Pr( \vec{N}_{1,\ast}  | \vec{d} ). \label{prob-d}
\end{eqnarray}
Now that we have derived $\Pr ( \vec{d} | \vec{u} )$, we express the conditional distribution of $\vec{\phi}$,
\begin{equation} \label{pst-dens}
 f ( \vec{\phi} | \vec{u} ) =  
  \sum_{\vec{d} \in {\cal D}} f ( \vec{\phi} ,\vec{d} |  \vec{u})
=  \sum_{\vec{d} \in {\cal D}} f ( \vec{\phi}  | \vec{d},   \vec{u}) \cdot \Pr(  \vec{d} | \vec{u}),
 \end{equation}
where $f ( \vec{\phi} |  \vec{d} , \vec{u} )$ is the conjugate posterior density of $\vec{\phi}$ given $\vec{d}$ and $\vec{u}$  for the hBeta model.

\begin{exa} \label{exa112}
According to Expression  (\ref{eq671}), the conditional segmentation probability, $\Pr( \vec{d} | \vec{u})$, is proportional 
to the product of the reciprocal of the multinomial coefficient for the counts vector $\vec{N}_{L, \ast}$  in (\ref{ inv-mult-coef}) 
and the conditional probability of the counts vector, $Pr( \vec{N}_{L, \ast}   | \vec{d})$, which is determined by $a_0$.

Recall that  in Model \ref{gen:model} given $\vec{\pi}_L$ and $\vec{d}$, $\vec{N}_{L, \ast}$ is multinomial with sample size $m$ and probabilities vector $\vec{\pi}_L$.
Thus the results of Example \ref{exa111} suggest that for  small values of $a_0$ large probabilities will be given to unevenly distributed counts vectors
and that increasing $a_0$ will favour  more evenly distributed counts vectors.

\medskip
To illustrate the effect of $a_0$ on $\Pr( \vec{d} | \vec{u})$, in Table 1
we list conditional segmentation probabilities for a set of $L = 2$ level  segmentations, ${\cal D} = \{ \vec{d}_1, \cdots \vec{d}_5 \}$, 
and a sample of $m=4$ observations,  $\vec{u} = (u_1, \cdots, u_4)$, in the  unit cube $[0,1]^P$.
The rows of Table 1 correspond to the segmentations, $\vec{d}_1 \cdots \vec{d}_5$.
The counts vector, $\vec{N}_{2, \ast} ( \vec{u} ; \vec{d}_j) = ( N_{2,1}, \cdots, N_{2,4})$, for $\vec{d}_j$ is listed in Column 1.
In columns 2-4, we list  $\Pr(  \vec{d}_j | \vec{u} )$ for $a_0 = 0.1, 1, 10$.

Indeed, we see that increasing $a_0$ provides larger probabilities to the segmentations that yield more evenly distributed count vectors at the first rows of Table 1.
Furthermore, comparing the second and third rows reveals that the segmentation probabilities is not exchangeable in $\vec{N}_{2, \ast}$,
where for each value of $a_0$, segmentations that yield contiguous non-zero counts are more probable.
Yet, the two segmentations yielding counts vector with a single non-zero entry are equally probable.
\end{exa}

\begin{table}[h!]
\centering
\begin{tabular}{c|| c |c |c|} 
 
 $\vec{N}_{2, \ast} (\vec{u} ; \vec{d}_j)  $ & $a_0 = 0.1$ & $a_0 = 1$ & $a_0 = 10$ \\ [0.5ex] 
 \hline 
 $(1, 1, 1, 1)$ & 0.00 & 0.01 & 0.13 \\ 
 $(0, 2, 0, 2)$ & 0.00 & 0.04 & 0.16 \\
 $(0, 0, 2, 2)$ & 0.01 & 0.07 & 0.19 \\
$(0, 0, 0, 4)$ & 0.49 & 0.44 & 0.26  \\ 
$(0, 0, 4, 0)$ & 0.49 & 0.44 & 0.26  \\ [1ex] 
\end{tabular}
\caption{Conditional segmentation probabilities.}
\label{table:1}
\end{table}

 
\bigskip
\subsection{Predictive inference}
The predictive distribution is the distribution of $u_{m+1}$ for generative model \ref{gen:model}.
Let $f (u)$ denote the density function for the predictive distribution,
which is a mixture of  $f ( u | \vec{d}, \vec{\phi} )$ for random $\vec{d} \in {\cal D}$ 
and $\vec{\phi}$.
The predictive probability of  ${\cal U} \subset [0,1]^P$ is given by
\begin{equation} \label{exp89}
\Pr( u_{m+1} \in {\cal U})  =  \int_{u \in {\cal U}}  f(u)  \  du.
\end{equation}

Let ${\cal U'}$ be a sufficiently small subset, so that $\forall \vec{d} \in {\cal D}$ it is a subset of a single level $L$ subinterval.
As  $f ( u | \vec{d}, \vec{\phi} )$ is constant on ${\cal U'}$ for all $ \vec{d}$, then $f(u)$ is constant on ${\cal U'}$ 
and therefore 
\[
\Pr( u_{m+1} \in {\cal U'})  =  volume({\cal U'}) \cdot  f(u).
\]
And in general, $f(u)$ is piecewise constant.

For $l = 1 \cdots L$,  let $j' (l, \vec{d})$ denote the index of the $l$ level subinterval for segmentation $\vec{d} \in {\cal D}$ that covers ${\cal U'}$.
Per construction $\forall \vec{d} \in {\cal D}$ and $\forall u \in {\cal I}_{L, j' (L, \vec{d})} (\vec{d})$,
\[  
f ( u |\vec{d}, \vec{\phi} ) =  2^L \cdot \pi_{L, j' (L, \vec{d})} (\vec{d}),
\]
for  $\pi_{L, j' (L, \vec{d})}$ that is a product of  $\phi_{1, j' (1, \vec{d})}  \cdots \phi_{L, j' (L, \vec{d})} $.
Therefore given $\vec{d}$ and $\vec{\phi}$, we may express the conditional predictive probability of ${\cal U'}$,
\begin{equation}
\Pr( u_{m+1} \in {\cal U'} |  \vec{d}, \vec{\phi} ) 
 = volume({\cal U'}) \cdot   2^L \cdot \Pi_{l = 1}^L  \phi_{j, j' (j, \vec{d})} (\vec{d}).    \label{pred-prob11}
\end{equation}

\medskip
As any Bayesian model, our inferential framework provides Bayesian estimates and credible regions for the model parameters.
For predictive inference we further consider three types of outcomes:

\medskip
\noindent {\bf a. Predictive samples.} Predictive samples, $u^1_{m+1} \cdots u^n_{m+1}$, are iid samples from the predictive distribution.
Predictive samples  may be generated by repeating the following steps for $i = 1 \cdots n$:
 Sample $( \vec{d}, \vec{\phi} (\vec{d}))$, compute $\vec{\pi}_L  (\vec{d})$,  sample $u^i_{m+1}  \sim  f ( u |\vec{d}, \vec{\pi}_L (\vec{d})  )$.
 
\medskip
\noindent {\bf b. Predictive probabilities.} The predictive probability of ${\cal U}$ is 
$\Pr( u_{m+1} \in {\cal U})$  for generative model \ref{gen:model}.
This outcome may be numerically evaluated by computing the proportion of predictive samples in ${\cal U}$,
\[  \Pr( u_{m+1} \in {\cal U})  \;  \widehat{=}  \; \; \frac{ |( i : u^i_{m+1} \in {\cal U} ) |}{  n}.
\]
In the next two subsections we analytically derive the predictive probability of ${\cal U'}$.

\medskip
\noindent {\bf c. Credible prediction sets.}  A level $1 - \alpha$ credible prediction  set is ${\cal U}_{1 - \alpha} \subset [0,1]^P$
with predictive probability $1 - \alpha$. Thus the number predictive samples in ${\cal U}_{1 - \alpha}$ is a $Binomial (n, 1-\alpha)$ random variable.


\bigskip
\subsubsection{The prior predictive distribution}
The prior predictive distribution is the marginal distribution of $u_{m+1}$  in generative model  \ref{gen:model},
in which $\vec{d}$ is sampled from ${\cal D}$ with equal probabilities and $\phi_{l, j} (\vec{d})$ are iid $Beta (a_0, a_0)$.
Thus using  (\ref{pred-prob11}) we may express the predictive probability of ${\cal U'}$, 
\begin{eqnarray} 
\lefteqn{ \Pr \left(  u_{m+1} \in {\cal U'} \right) = E_{\vec{\phi} , \vec{d}}  \left[ \ \Pr \left(  u_{m+1} \in {\cal U'} | \vec{d} , \vec{\phi}  \ \right) \right] 
 = E_{\vec{\phi},  \vec{d}}  \left[ volume({\cal U'}) \cdot   2^L \cdot \Pi_{l = 1}^L  \phi_{j, j' (l, \vec{d})} (\vec{d}) \right]  }   \nonumber \\
& =  & volume({\cal U'}) \cdot   2^L \cdot     E_{ \vec{d}} \  \Pi_{l = 1}^L  E_{\vec{\phi} | \vec{d}}  \left[    \phi_{l, j' (j, \vec{d})} (\vec{d}) \right]  
= volume({\cal U'}) \cdot   2^L \cdot  \Pi_{l = 1}^L  \left( \frac{a_0}{ a_0 + a_0} \right) \nonumber \\
&   = &  \  volume({\cal U'}).  \label{exp56}  
 \end{eqnarray}
 Therefore the prior predictive distribution is the uniform distribution with $f(u) \equiv 1$.

\subsubsection{The posterior predictive distribution} \label{sec:pdd}
The posterior predictive distribution of $u_{m+1}$ is the conditional distribution of  $u_{m+1}$ given 
$\vec{u} = (u_1 \cdots u_m)$  in generative model  \ref{gen:model}.

\medskip
Conditional on $\vec{d}$ and on $\vec{N} = \vec{N}_{L, \ast}  (\vec{u} ; \vec{d})$, 
 $ \pi_{L, j' (L,\vec{d})} = \Pi_{l = 1}^L \phi_{l, j' (l, \vec{d})}$,
with   independent $\phi_{1, j' (1, \vec{d})} \sim Beta( a_0 + N_{1,j'(1 ,\vec{d})}, a_0 + m  - N_{1,j'(1 ,\vec{d})} )$ 
and $\phi_{l, j' (l, \vec{d})} \sim Beta( a_0 + N_{l,j'(l ,\vec{d})}, a_0 + N_{l-1,j'(l-1 ,\vec{d})} - N_{l,j'(l ,\vec{d})} )$ for $l = 2 \cdots L$.
Thus using  (\ref{pred-prob11}) we may express
\begin{eqnarray} 
\lefteqn{ \Pr \left(  u_{m+1} \in {\cal U'}  |  \vec{N}, \vec{d}  \ \right) = E_{\vec{\phi} | \vec{N}, \vec{d}}  \left[ \ \Pr \left(  u_{m+1} \in {\cal U'} | \vec{d} , \vec{\phi}  \ \right) \right] 
  }   \nonumber \\
& =  & volume({\cal U'}) \cdot   2^L \cdot  \Pi_{l = 1}^L  E_{\vec{\phi} | \vec{N}, \vec{d}}  \left[    \phi_{l, j' (l, \vec{d})} (\vec{d}) \right]  
 \nonumber \\
&   = &   volume({\cal U'}) \cdot   2^L \cdot   
\left[ \Pi_{l = 1}^L  \left( \frac{ N_{l, j' (l , \vec{d})} + a_0}{N_{l-1, j' (l-1 , \vec{d})}  + 2 \cdot  a_0} \right) \right],   \label{exp56a}  
 \end{eqnarray}
with $N_{0, j' (0 , \vec{d})}  = m$. 
Recalling that the sequence subinterval counts is  decreasing,  $m \ge  N_{1, j' (1, \vec{d})} \ge  \cdots \ge  N_{L, j' (L, \vec{d})}$,
let $L'$ denote the largest $l$ with a nonzero count, i.e. $N_{L', j' (L', \vec{d})} > 0$. 
Therefore we may rewrite Expression (\ref{exp56a}),
\begin{eqnarray} 
\lefteqn{
 \Pr  \left(  u_{m+1} \in{\cal U'} | \vec{N}, \vec{d} \   \right)  } \nonumber \\
  & =  &  volume({\cal U'}) \cdot   2^L \cdot  
\frac{ N_{1, j' (1, \vec{d})} + a_0}  {m + 2 a_0} 
\cdot  
\frac{ N_{2, j' (2, \vec{d})} + a_0}  {N_{1, j' (1, \vec{d})} + 2 a_0} 
\cdot  \; \cdots \; 
\cdot  
\frac{ N_{L', j' (L', \vec{d})} + a_0}  {N_{L'-1, j' (L'-1, \vec{d})} + 2 a_0}  \cdot   \left( \frac{ a_0 }{2 a_0} \right)^{L - L'} \nonumber \\ 
  & =  & volume({\cal U'})   \cdot  2^{L'}  \cdot 
\frac{ N_{1, j' (1, \vec{d})} + a_0}  {m + 2 a_0} 
\cdot  
\frac{ N_{2, j' (2, \vec{d})} + a_0}  {N_{1, j' (1, \vec{d})} + 2 a_0} 
\cdot  \; \cdots \; 
\cdot  
\frac{ N_{L', j' (L', \vec{d})} + a_0}  {N_{L'-1, j' (L'-1, \vec{d})} + 2 a_0},  \label{exp22c}
\end{eqnarray}
and thus for $u \in {\cal U'}$ the conditional posterior predictive density is 
\begin{equation} \label{eq45}
f( u | \vec{N}, \vec{d}) = 
 2^{L'}  \cdot 
\frac{ N_{1, j' (1, \vec{d})} + a_0}  {m + 2 a_0} 
\cdot  
\frac{ N_{2, j' (2, \vec{d})} + a_0}  {N_{1, j' (1, \vec{d})} + 2 a_0} 
\cdot  \; \cdots \; 
\cdot  
\frac{ N_{L', j' (L', \vec{d})} + a_0}  {N_{L'-1, j' (L'-1, \vec{d})} + 2 a_0}.
\end{equation}

Note that for  $a_0 \rightarrow 0$, the numerator in the $l$'th term and the denominator of the
$l+1$'th term in the product in (\ref{eq45}) are equal for $l = 1 \cdots L'-1$,
thereby yielding 
\begin{equation} \label{lim11}
\lim_{a_0 \rightarrow 0} f( u | \vec{N}, \vec{d}) 
=    2^{L'}     \cdot    \frac{N_{L', j' (L', \vec{d})}}{ m}.
\end{equation}
On the other hand, increasing $a_0$ shrinks  each of the $L'$ fraction terms in  (\ref{eq45}) to $1/2$, yielding
 \begin{equation} \label{lim11a}
\lim_{a_0 \rightarrow \infty} f( u |  \vec{N}, \vec{d}) 
=    2^{L'}     \cdot    (1/2)^{-L'} = 1.
\end{equation}

\medskip
Lastly, the posterior predictive probability of ${\cal U'}$ is a mixture of the 
conditional posterior predictive probabilities in (\ref{exp22c}),
\begin{equation}
 \Pr  \left(  u_{m+1} \in{\cal U'} | \vec{u}  \right)   = \sum_{\vec{d} \in {\cal D}}  \Pr  \left(  u_{m+1} \in{\cal U'}, \vec{d}  | \vec{u}  \right)
 = \sum_{\vec{d} \in {\cal D}}  \Pr  \left(  u_{m+1} \in{\cal U'}  |     \vec{N} (\vec{u}; \vec{d}), \vec{d} \ \right) \cdot  \Pr (  \vec{d}  | \vec{u} ),
\end{equation}
where $\Pr (  \vec{d}  | \vec{u} )$ is the posterior segmentation probability in (\ref{eq671}).
And the posterior predictive density for $u \in {\cal U'}$  is a mixture of the conditional densities in (\ref{eq45}),
\[
f(u | \vec{u})  =    \frac{  \Pr  (  u_{m+1} \in{\cal U'} | \vec{u} )}{ volume( {\cal U'})}  
 =  \sum_{\vec{d} \in {\cal D}}  f  ( u  |     \vec{N} (\vec{u}; \vec{d}), \vec{d}  \ ) \cdot  \Pr (  \vec{d}  | \vec{u} ).
\]

\subsection{Frequentist properties of our inferential framework}
Recall, the underlying assumption in this report is that the real data generative model is  that $u_1 \cdots u_{m+1}$ are iid $\tilde{\pi}$.
We will use subscript $\tilde{\pi}$ to denote real probabilities,
all other probability statements  are with respect to model 
\ref{gen:model}.

\medskip
We begin by considering a single segmentation $\vec{d}$.
Let $\tilde{\pi}_L (\vec{d})  = (\tilde{\pi}_{L,1}, \cdots, \tilde{\pi}_{L, 2^L})$ denote the vector of probabilities 
that $\tilde{\pi}$ assigns to subintervals  ${\cal I}_{L,1} (\vec{d}), \cdots, {\cal I}_{L,2^L} (\vec{d})$.
Thus the real sampling distribution of $\vec{N} = \vec{N}_{L, \ast}  (\vec{u} ; \vec{d})$ is Multinomial with number of trials $m$ and event probabilities vector $\tilde{\pi}_L (\vec{d}) $.
In our inferential framework,
the real probability that $u_{m+1} \in {\cal U'}$ is estimated by the conditional posterior predictive probability
\[
\Pr_{\tilde{\pi}} ( u_{m+1} \in {\cal U'}) \;  \widehat{=}  \; \Pr  \left(  u_{m+1} \in{\cal U'} | \vec{N}, \vec{d} \   \right)
= volume ( {\cal U'}) \cdot f(u | \vec{d}, \vec{N}).
\]

To assess how well we evaluate $\tilde{\pi}$ with a given segmentation $\vec{d}$, we consider two types of errors:
(a) approximation error, defined as the distance between $\tilde{\pi}$ and step-function density 
\begin{equation} \label{oracle_dens}
f ( u ; \vec{d}, \tilde{\pi}_L )  =  \tilde{\pi}_{L,1} \cdot  \frac{ I_{L,1} (u  ; \vec{d}) }{(1/2)^L} +  \cdots +
\tilde{\pi}_{L, 2^L} \cdot \frac{ I_{L, 2^L} (u  ; \vec{d}) }{ (1/2)^L},
\end{equation}
which is an irreducible term that depends on the smoothness of $\tilde{\pi}$ with respect to $\vec{d}$;
(b) estimation error of $f ( u ; \vec{d}, \tilde{\pi}_L )$ by $f ( u | \vec{d}, \vec{N})$,
which is reduced by increasing $m$.
The estimation error may be further divided into  bias and variance terms.

In general, coarse segmentations have large approximation error and small estimation error.
Refining the segmentation by increasing $L$,  decreases the approximation error and increases the estimation error.
For coarse segmentation $\vec{d}$ and sufficiently large $m$  that ensures that $N_{L, j'(L,\vec{d'})} >0$ and thus $L' = L$,
Expression (\ref{lim11}) implies that for $u \in {\cal U'}$,
\begin{eqnarray*}
E_{u_1 \cdots u_m \sim \tilde{\pi}} 
\lim_{a_0 \rightarrow 0} f( u | \vec{d}, \vec{N}) 
& =  &   E_{ \tilde{\pi}}  \frac{ 2^{L}  \cdot   N_{L, j'(L,\vec{d'})} (\vec{d}) }{m}     =    \frac{ E_{ \tilde{\pi}}   N_{L, j'(L,\vec{d'})} (\vec{d}) }{ (1/2)^L \cdot m}  \\
& = &  \frac{ m \cdot \tilde{\pi}_{L, j' (L, \vec{d'})}}{ (1/2)^L \cdot m} = f(u ; \vec{d}, \tilde{\pi}_L).
\end{eqnarray*}
Therefore for small $a_0$ and coarse segmentations 
 $f ( u | \vec{d}, \vec{N})$ has small bias.
 
On the other hand,  if $L' < L$ then according to Expression (\ref{eq45}), $f(u | \vec{d}, \vec{N}) = f(u | \vec{d''}, \vec{N})$ for 
any segmentation $\vec{d''}$ that is a refinement of $\vec{d}$. As a result, our density estimators maintain small estimation error
for fine segmentations. 
This is especially useful for larger values of $a_0$ that  yield biased density estimation with small estimation variance.
According to the simulations in Subsection 3.1, setting $a_0 = 1$ provide small estimation error even for fine segmentations and small $m$.

\medskip
Our inferential framework considers multiple segmentations $\vec{d}$, with different approximation and estimation errors,
that are weighted according to the posterior segmentation probability $\Pr( \vec{d} | \vec{u})$.
The most interesting and surprising feature of our framework is that for $a_0 = 1$ 
segmentations with large posterior probabilities tend to have small approximation and estimation errors.  
We will demonstrate this in simulations in Subsection 3.2.

Our inferential framework also produces  level $1 - \alpha$ credible prediction sets, ${\cal U}_{1 - \alpha} \subset [0,1]^P$.
Per construction, credible prediction sets have posterior predictive probability $1 - \alpha$, $\Pr( u_{m+1} \in  {\cal U}_{1 - \alpha} | \vec{u}) = 1 - \alpha$.
In Section 4, we will demonstrate how our inferential framework may be adapted to produce conformal prediction sets 
 $\tilde{\cal U}_{1 - \alpha} \subset [0,1]^P$, for which $\Pr_{\tilde{\pi}} ( u_{m+1} \in \tilde{\cal U}_{1 - \alpha} ) \ge 1 - \alpha$.

\bigskip
\section{Density estimation simulations}

\subsection{Simulation study of 1D density estimation}
In this subsection we present the results of simulations studying the performance of our density estimators for estimating $\tilde{\pi}$ that has a piecewise continuous density on $[0,1]$.
In each simulation run we generate a sample of  iid observations $\vec{u} = (u_1, \cdots, u_m)$ from $\tilde{\pi}$ and use it 
 to compute $f ( u | \vec{d}, \vec{N})$ for $L = 10, 5, 3$, and a set value of $a_0$.
As we estimate a $P = 1$ dimensional density, we consider the canonical segmentation  of ${\cal I}_0 = [0,1]$, in which 
${\cal I}_{l,j}  = [(j-1) \cdot 2^{-l}, j \cdot 2^{-l}]$ for $l = 1 \cdots L$ and $j = 1 \cdots 2^l$.
We performed $4$ sets of simulations with $m = 50, 200$ and $a_0 = 1, 0.1$, each consisting of $10,000$ simulation runs. 

\medskip
The simulation results are presented in Figure 4. Each pair of plots corresponds to the same set of simulations:
the left plot displays the simulation mean of  $f ( u |\vec{d}, \vec{N} )$; 
the right plot displays the square root of the simulation mean of the estimator squared-error, 
$ \left(  \hat{\pi} (u)  - \tilde{\pi}  ( u ) \right)^2$,
for  $\hat{\pi} (u) = f ( u |\vec{d}, \vec{N} )$ and for the 
interval-counts density estimators, $ \hat{\pi} (u) =  2^L \cdot N_{L, j} (\vec{u} ; \vec{d}) / m$.
The black curve in all the plots is $\tilde{\pi} (u)$,  the piecewise continuous density we are estimating.
The green, blue, red curves correspond to level $L=10, 5, 3$, segmentations of $[0,1]$.
Solid curves are drawn for the hBeta estimators and dashed curves are drawn for the  interval-counts estimators.

\medskip
The estimator mean plots reveal estimation bias and approximation error.
For $m = 50$ and $a_0 = 1$ the hBeta density estimators are biased 
-- upward bias for small density values 
and downward bias for large density values.
For $m = 200$ and $a_0 = 1$ the bias of the estimators is smaller.
For $a_0 = 0.1$ the bias is considerably smaller and can only be observed for the small density values.
The bias does seem to depend on $L$.

To assess approximation errors, which do not depend on $m$ and $a_0$,  we focus on the $m = 200$ and $a_0 = 0.1$ configuration that has the smallest bias.
There is no approximation error in the regions in which  $\tilde{\pi} (u)$ is constant.
In the region in which $\tilde{\pi} (u)$ is linear in $u$, there is considerable approximation for $L=3$ that decreases as $L$ increases.

\medskip
The estimator MSE is the sum of the  squared estimation error and the squared approximation error.
In this example the estimation error is considerably larger than the approximation error.
The hBeta estimator MSE decreases in $m$, and given $m$ and $a_0$ it increases in $L$.
For $L = 3$, $m = 200$ and $a_0 = 0.1$, the MSE of the hBeta estimator is the same as the MSE
of the interval-counts estimator, which is unbiased with $MSE = (2^L)^2  \cdot \tilde{\pi} (u) \cdot (1 - \tilde{\pi} (u)) / m$ for all $L$ and $m$.
The ratio between the MSE of the  interval-counts estimator and the hBeta estimator increases for larger $L$ and $a_0$ and smaller $m$.
For $L = 10$, $m = 50$ and $a_0 = 1$, the sqrt(MSE) of the hBeta estimator is $4-5$ times smaller than that of the interval-counts estimator.

 \begin{figure}[h] 
\centering
\includegraphics[height=.49\textwidth]{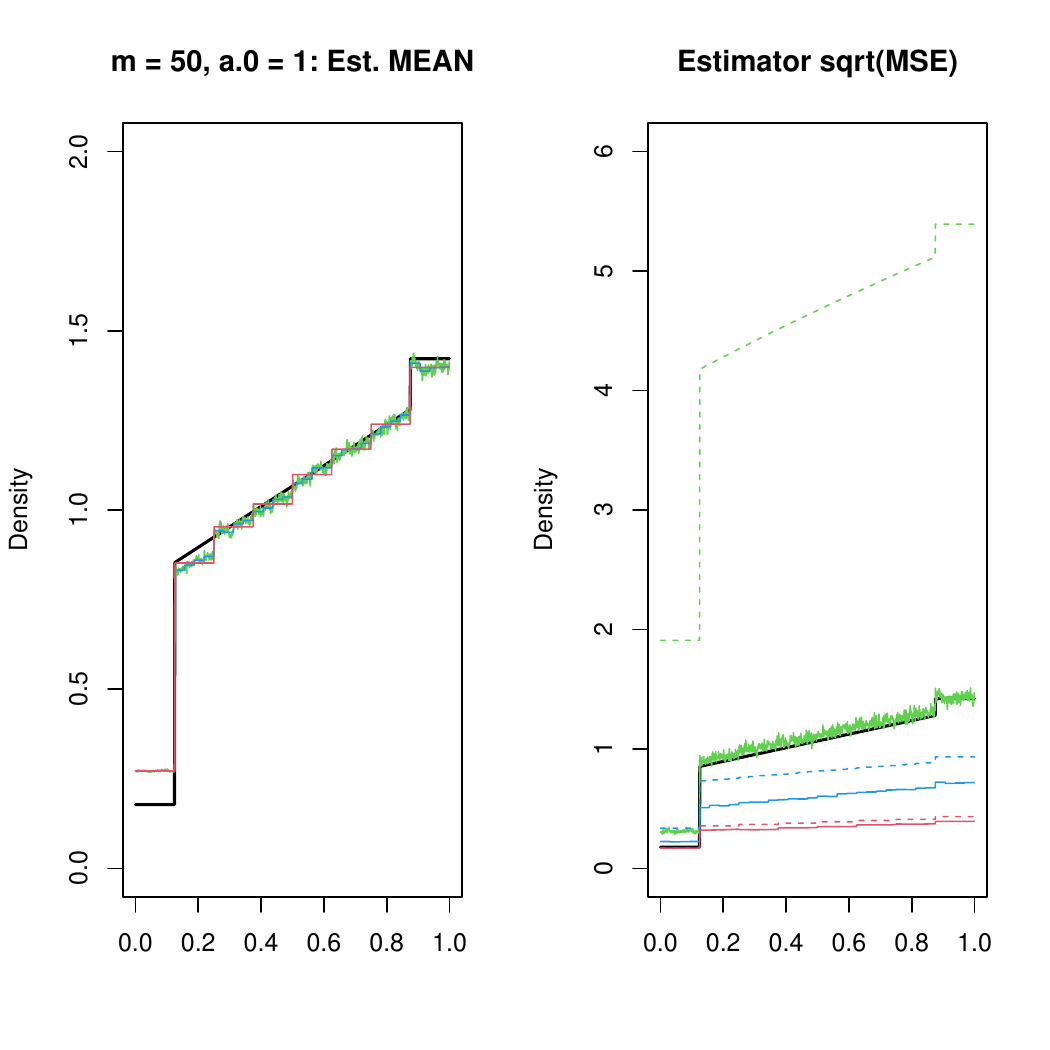}
\includegraphics[height=.49\textwidth]{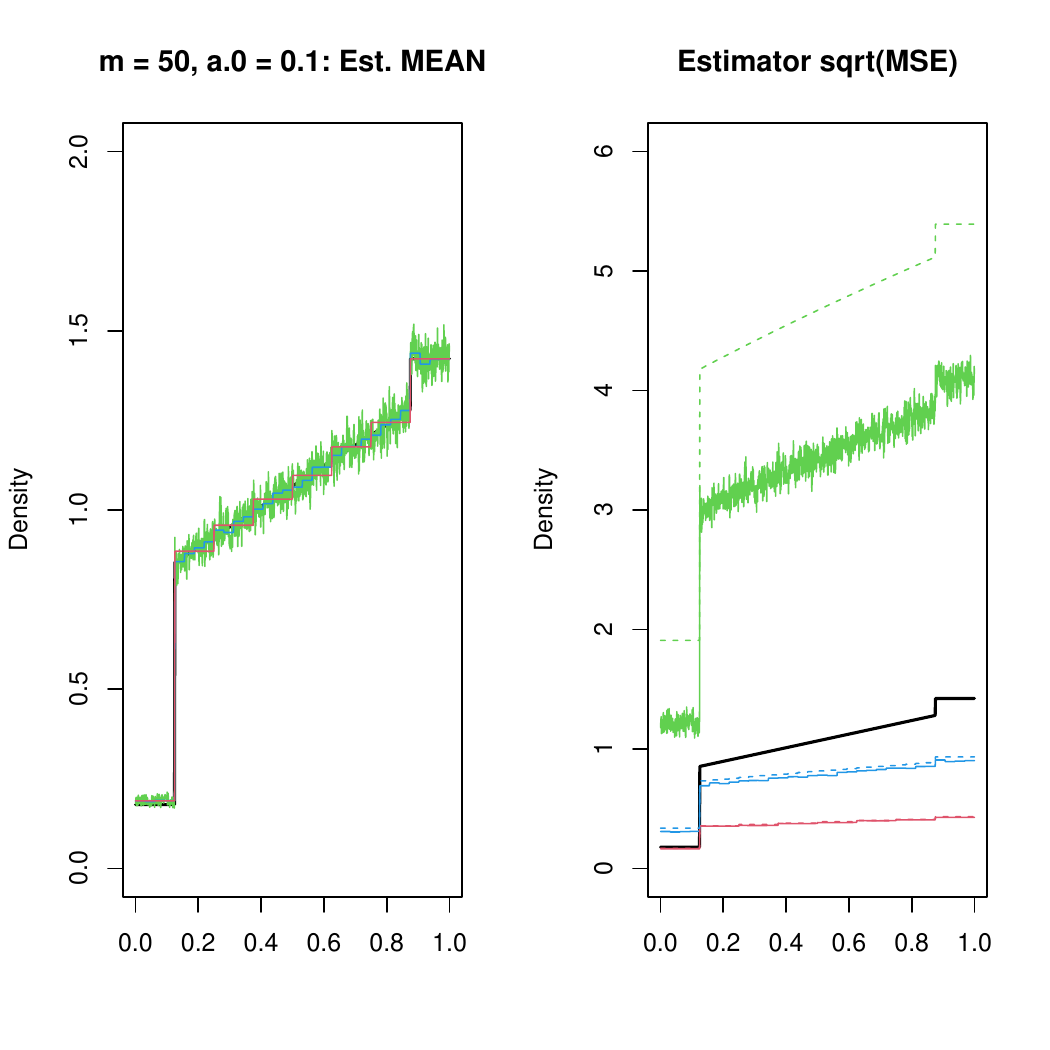}
\includegraphics[height=.49\textwidth]{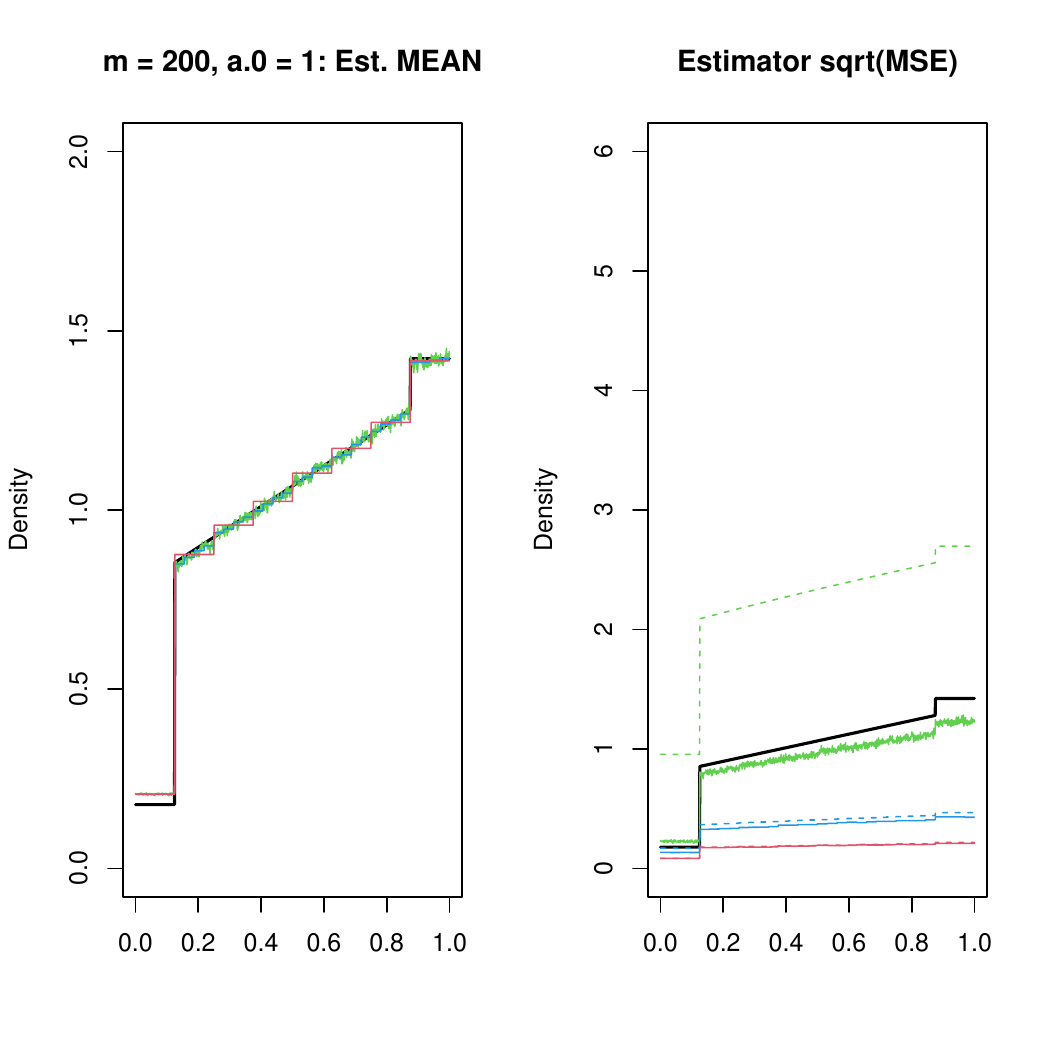}
\includegraphics[height=.49\textwidth]{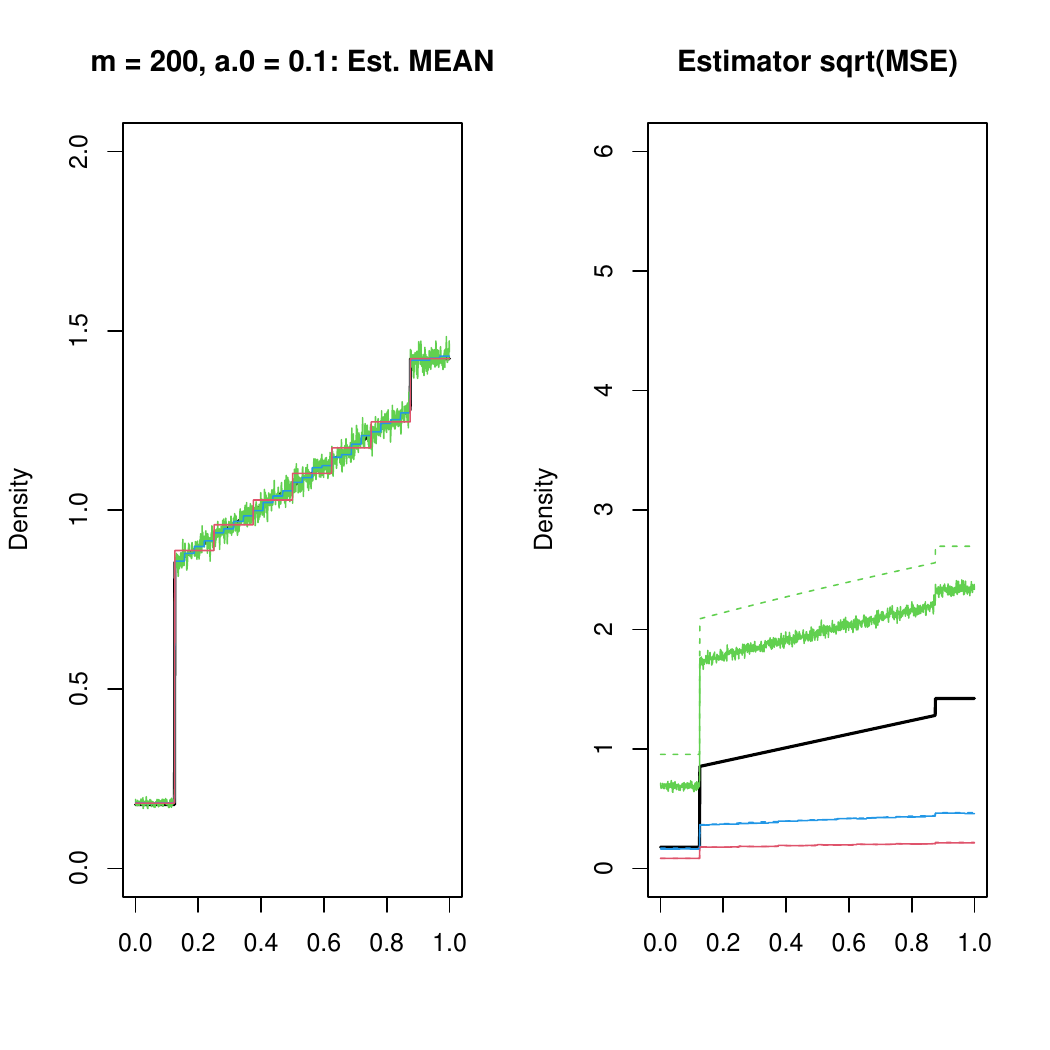}
\caption{1D density estimation -- 
results of $4$ simulations with $m = 50, 200$ and $a_0 = 1, 0.1$.
Each pair of plots corresponds to the same simulations:
the left plot is the simulation mean of the estimators;
the right plot is the square root of the simulation MSE of the estimators.
The black curve in all the plots is  $\tilde{\pi} (u)$.
The green,  blue, red  curves correspond to $L=10, 5, 3$ level segmentation of $[0,1]$. 
The solid curves correspond to hBeta estimators;
the dashed curves are the square-root MSE for the interval-counts density estimators.
}
\end{figure}

 \bigskip
 \subsection{Simulation study of 2D density estimation}
In this subsection, for $u = (u_x, u_y)$,
we present simulation results for
estimating $\tilde{\pi} ( u) = 2 \cdot logit^{-1} (20 \cdot (u_x-0.5))$
with $f( u | \vec{d}, \vec{N})$, for the four $L=4$ level segmentations of $[0,1]^2$ shown in Figure 1.

\medskip
The results  are displayed in Figure 5.
In the top  row of plots we show how well $\tilde{\pi}$ is approximated by $f (u ; \vec{d}, \tilde{\pi}_L)$ in the the four segmentations of $[0,1]^2$.
In each plot, the regions separated by the green vertical lines in each plot correspond to the $16$ subintervals of each segmentation,
${\cal I}_{4,1} (\vec{d}) \cdots {\cal I}_{4,16} (\vec{d})$, arranged from left to right.
Thus, in the four plots the last interval on the right is ${\cal I}_{4, 16} (\vec{d})$.
According to Figure 1, 
for segmentation $(X,X,X,X)$, ${\cal I}_{4, 16} (\vec{d})$  is the subinterval on the right end of $[0,1]^2$;
for segmentation $(Y,Y,Y,Y)$, ${\cal I}_{4, 16} (\vec{d})$,  is the  subinterval  on the top end of $[0,1]^2$;
for  segmentations $(X,X,Y,Y)$ and  $(Y,Y,X,X)$, ${\cal I}_{4, 16} (\vec{d})$ is the  subinterval  on the top right corner of $[0,1]^2$.

In the four plots,  the black curves display the profile of $\tilde{\pi} ( u)$ as a function of $u_x$ and the red horizontal lines display  $f (u ; \vec{d}, \tilde{\pi}_L)$,
 for each subinterval.
We summarize the approximation error for each segmentation by
the square root of a $MSE$ measuring the distance between $\tilde{\pi} ( u)$ and $f (u ; \vec{d}, \tilde{\pi}_L)$, 
\[
MSE \left( \tilde{\pi} ( u) , f (u ; \vec{d}, \tilde{\pi}_L) \right) :=
\int_{u \in [0,1]^2}  \left( \tilde{\pi} ( u)  -  f (u ; \vec{d}, \tilde{\pi}_L) \right)^2 du
 = \sum_{j = 1}^{16}  \int_{u \in {\cal I}_{4,j} (\vec{d})} \left( \tilde{\pi} ( u)  -   \tilde{\pi}_{L, j} (\vec{d})  \right)^2 du.
\]
Segmentation $(X,X,X,X)$ has the smallest approximation error, $sqrt(MSE) = 0.066$;
Segmentation $(Y,Y,Y,Y)$ has the largest approximation error,  $sqrt(MSE) = 0.894$;
the approximation error for segmentations $(X,X,Y,Y)$ and  $(Y,Y,X,X)$, 
that have the same set level $L=4$ subintervals, is  $sqrt(MSE) = 0.199$; 

\medskip
The simulation study consisted of $10,000$ simulation runs.
In each run, $m = 50$ iid $u_i \sim \tilde{\pi}(u)$ were generated and used to 
compute $\Pr( \vec{d} | \vec{u})$ and  the hBeta estimator, 
 $\hat{\pi}_{L,j}  (\vec{N} ; \vec{d})  = \Pr( u_{m+1} \in {\cal I}_{L,j} (\vec{d}) | \vec{N}, \vec{d})$, for each segmentation with $a_0 = 1$.
The estimation error of $\hat{\pi}_{L,j}  (\vec{N} ; \vec{d})$  is summarized by the Pearson residual,   
$ ( \hat{\pi}_{L,j}  (\vec{N} ; \vec{d})  - \tilde{\pi}_{L, j} (\vec{d})) / \sqrt{ \tilde{\pi}_{L, j} (\vec{d})}$.
Note that for each simulation run $\vec{N} \sim Multinomial (50 ; \tilde{\pi}_L (\vec{d}))$.
Therefore Pearson residual for the counts estimator, $\hat{\pi}_{L,j}  (\vec{N} ; \vec{d})  = N_{L, j} (\vec{d}) / 50$,
is approximately $N(0,1)$ and the resulting $X^2$ statistic, which is the sum of squares of the $16$ Pearson residuals, is a chi-square random variable with $15$ degrees of freedom.

The boxplots in the bottom row of Figure 5 display the distribution of the simulation outcomes.
The left and middle sets of boxplots display the distribution of $X^2$ 
and the distribution of the mean of the absolute value of the Pearson residuals,  for the four segmentations. 
The plots reveal that the estimation error for all four segmentations is considerably smaller than the estimation error
for the counts estimator, for which the expectation of the  $X^2$ statistic is $15$ and the expectation of the mean of the absolute value of the Pearson residual
is approximately  $E_{Z \sim N(0,1)} | Z | = 0.798$.
As our estimation algorithm shrinks the density at each level to the Uniform density, the estimation error is minimized for segmentation $(Y,Y,Y,Y)$
for which the $16$ estimated probabilities are equal, and increases as the differences in probabilities between contiguous subinterval increase.
Thus even though the set of estimated probabilities is the same for segmentations $(X,X,Y,Y)$ and  $(Y,Y,X,X)$,
the estimation error is much smaller for segmentation $(X,X,Y,Y)$ for which the set of probabilities is ordered.

The four  boxplots on the right display the distribution of posterior segmentation probability $\Pr( \vec{d} | \vec{u})$ for each segmentation.
The plots reveal that $(X,X,X,X)$ that has the smallest approximation error is the most probable segmentation.
Segmentation $(Y,Y,Y,Y)$ that has the largest approximation error has almost $0$ posterior probability.
Of the two segmentations with the same approximation error, segmentation $(X,X,Y,Y)$ that has the smaller estimation error has considerably larger 
posterior probabilities.

\begin{figure}[h] 
\centering
\includegraphics[width=1\textwidth,height=.5\textwidth]{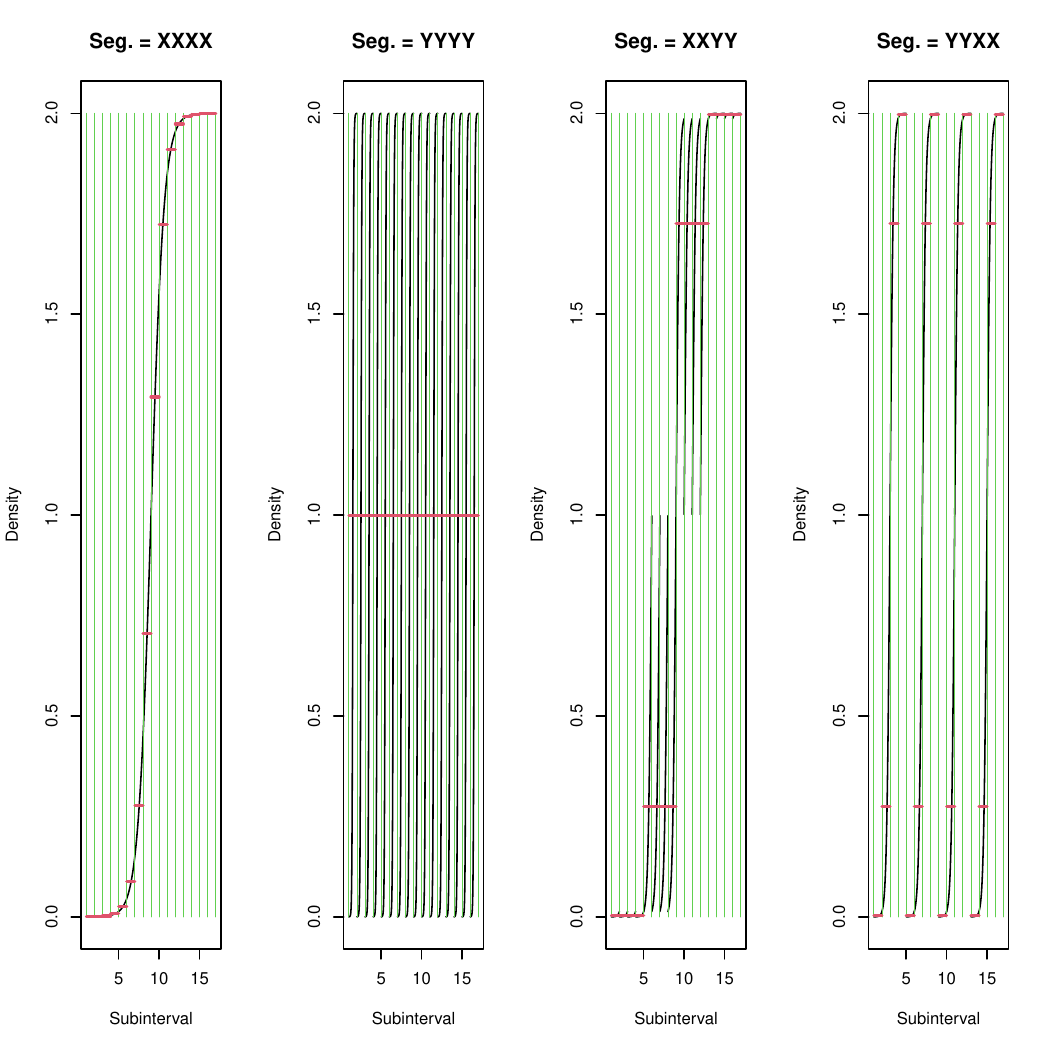}
\includegraphics[width=1\textwidth,height=.5\textwidth]{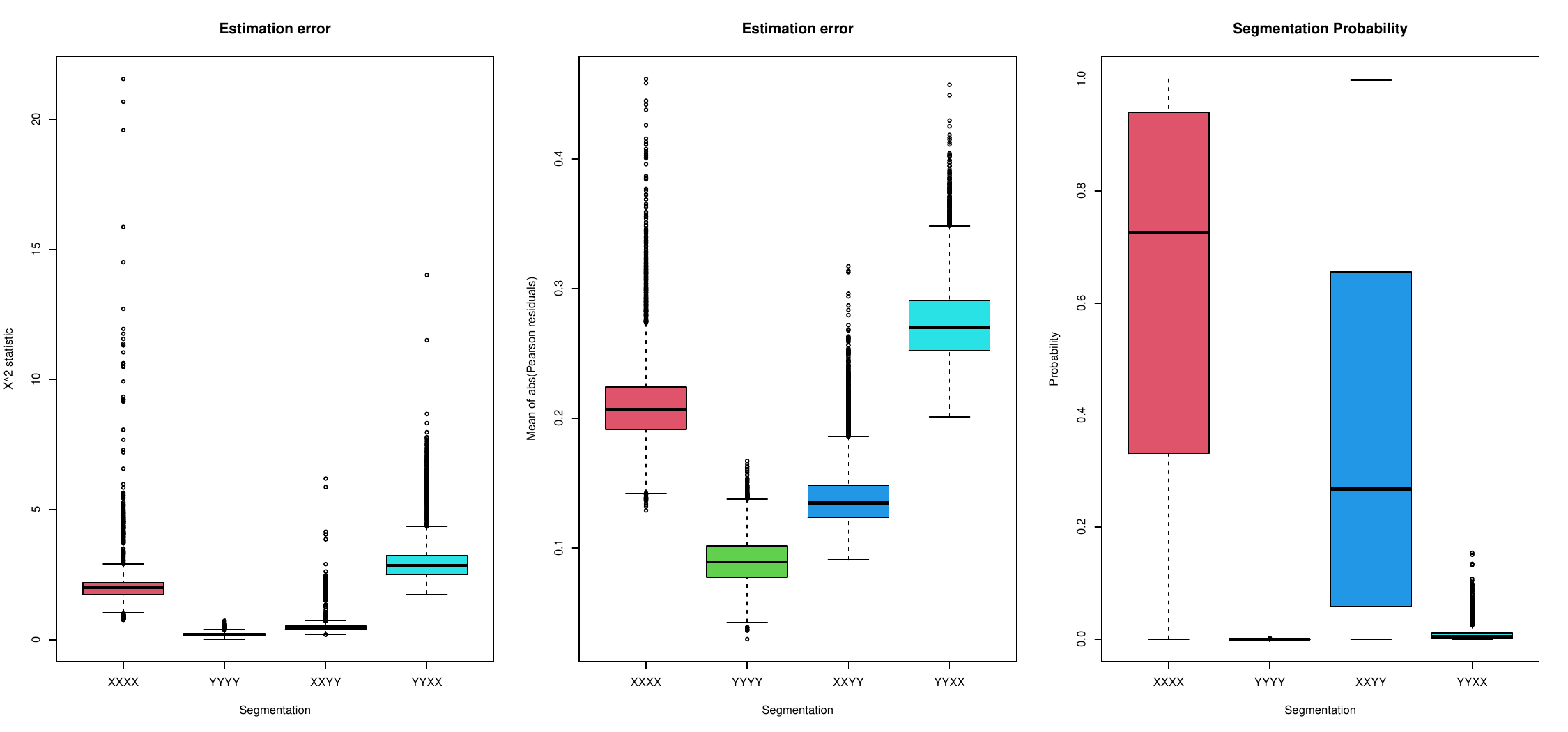}
\caption{2D density estimation  -- approximation error and simulation results for the four  segmentations of $[0,1]^2$ shown in Figure 1.
The plots on the top row  show how well $\tilde{\pi} ( u_x, u_y)$ is approximated by $f (u ; \vec{d}, \tilde{\pi}_L)$ for each segmentation.
In each plot, the regions between the green lines correspond to the $16$ subintervals of each segmentation;
the black curves display the profile of $\tilde{\pi} ( u_x, u_y)$ and the red lines display $f (u ; \vec{d}, \tilde{\pi}_L)$ in each subinterval of each segmentation.
The boxplots in the bottom row display simulation results.
The left and middle sets of boxplots display the distribution of $X^2$ 
and the distribution of the mean of the absolute value of the Pearson residuals, for the four segmentations. 
The four  boxplots on the right display the distribution of $\Pr( \vec{d} | \vec{u})$ for each segmentation.}
\end{figure}

\bigskip
\section{Quantile regression simulation}

In this simulated example $\tilde{\pi} (u)$ is the density of 
$U = (U_x, U_y)$, where $U_x = logit^{-1}(X)$ for  $X \sim N(0,4)$,
and $U_y = logit^{-1}(Y)$ with $Y | X  = x \sim N( -0.9, 0.25)$ for $x < -1$
 and $Y | X  = x \sim N( 0.9 \cdot x, 0.25)$  for $x \ge -1$.
We segment ${\cal I}_0 = [0,1]^2$ with
$L = 8$ dyadic segmentations. Similarly to the previous 2D example,
 we only consider $\vec{d}$ in which all the subintervals at the same level are partitioned according to the same dimension.
In this case ${\cal D}$ is the set of  $70 = { 8 \choose 4}$ segmentations consisting of four $X$ partitions and four $Y$ partitions 
that produce an array of $16 \times 16$ square subintervals,
and we set $a_0 = 1$. 

\medskip
The plots in Figure 6 correspond to simulations with sample size $m=100$ (top) and $m = 1000$ (bottom).
In each plot, 
the red X's mark the observations, $u_1 \cdots u_m$,
and the blue curves are the $0.05$, $0.50$ and $0.95$ quantiles of  $\tilde{\pi} ( U_y | U_x = u_x)$. 
Per construction, for all $u_x \in [0,1]$ the interval between the dashed blue lines forms a $0.90$ conditional predictive interval for $U_y  \sim \tilde{\pi} ( U_y | U_x = u_x)$.
Thus the region between the dashed blue lines is a $0.90$ predictive set for $U  \sim \tilde{\pi}$.

In order to produce predictive inferences, for each segmentation $\vec{d}_j \in {\cal D}$,
we compute the posterior segmentation probability $\Pr( \vec{d}_j | \vec{u})$,
 and generate $50$  Beta vectors from the conditional distribution of $\vec{\phi}$ given $\vec{d}_j$ and $\vec{N} ( \vec{u}  ; \vec{d}_j)$
 and use them to compute probability vectors, 
$\vec{\pi}^{j,1}_L \cdots \vec{\pi}^{j,50}_L$.
We approximate the posterior predictive density $f(u | \vec{u})$ by the mixture density
\begin{equation} \label{mix3}
\sum_{j = 1}^{70} \sum_{h = 1}^{50}   \frac{\Pr( \vec{d}_j | \vec{u} )}{50} f( u | \vec{\pi}^{j,h}_L, \vec{d}_j).
\end{equation}
The Green circles in each plot, are $2000$ posterior predictive samples, $u^1_{m+1} \cdots u^{2000}_{m+1}$, drawn independently from mixture density (\ref{mix3}).
The Orange curves are the $0.05$, $0.50$ and $0.95$ quantiles of the conditional distribution of $U_y$ given $U_x = u_x$ 
for mixture density (\ref{mix3}). Note that as the mixture density is piecewise constant on an array of $16 \times 16$  subintervals,
each quantile profile is piecewise constant on the $16$ subintervals of $U_x$.
For all $u_x \in [0,1]$ the interval between the dashed Orange lines form a $0.90$ conditional posterior predictive interval for $U_y | U_x = u_x$,
thus the region between the dashed lines is a $0.90$ posterior credible prediction set.

The plots reveal that  the posterior predictive density captures the general shape of $\tilde{\pi}$ pretty well.
For  $m = 100$, the  conditional posterior predictive distribution is more spread out  than $\tilde{\pi} ( U_y | U_x = u_x)$,
especially for values of $u_x$ close to $1$, for which  $\tilde{\pi} ( U_y | U_x = u_x)$ converges to $1$.
For $m=1000$,  the distribution of the posterior predictive samples is more similar to the distribution of $u_1 \cdots u_m$,
and other than an approximation error,
the quantiles of the conditional posterior predictive distribution and the quantiles of $\tilde{\pi} ( U_y | U_x = u_x)$ are almost the  same for all $u_x$.

\subsection{Conformal Prediction sets}
Vovk et al. (2005) present a general algorithm for constructing prediction sets for the following setting:
$w_1 \cdots w_{m+1}$ are iid $\tilde{\pi} (w)$, with $w_i = (x_i, y_i)$.
On observing the training set $\{ w_1 \cdots w_{m} \}$ and $x_{m+1}$, the goal is to construct a prediction set for $y_{m+1}$.
The key component for constructing the prediction sets is the conformity score, $A( \{ w_1 \cdots w_{m} \}, w) \in \mathbb{R}$,
that shows how well an additional point $w = (x, y)$ conforms to the training set.
They prove that Conformal Prediction Sets constructed in Algorithm \ref{dfn56} 
cover $y_{m+1}$ with probability $\ge 1 - \alpha$.

\begin{dfn} {\bf Algorithm for constructing Conformal Prediction Sets} \label{dfn56}
\begin{enumerate}
\item	For each potential  value $y$ of $y_{m+1}$, compute the conformity score
\[
 	a_{m+1} (y) =   A \left(  \{ w_1, \cdots,   w_m \} , (x_{m+1}, y) \right).
\]
To assess the significance of $a_{m+1} (y) $,  for $i = 1 \cdots m$ compute conformity score 
\[
a_i  (y) = A  \left(  \{ w_1, \cdots,  w_{i-1}, (x_{m+1}, y), w_{i+1}, \cdots, w_m \} , w_i \right),
\]
and compute the rank-based p-value 
\[  
p(y) = \frac{  | \{ i  : a_i (y)  \le a_{m+1} (y) \} |}{m+1}.
\]
\item The $1 - \alpha$ Conformal Prediction Set for $y_{m+1}$ is then defined
\[
\tilde{\cal Y}_{1 - \alpha} := \{ y: p(y) > \alpha \}.
\]
\end{enumerate}
\end{dfn}

\medskip
To construct conformal prediction sets for the quantile regression example,
we define the conformity score between training set $\vec{u}$ and an additional point $u_{m+1} = (u_x, u_y)$  to be 
the conditional posterior predictive given $\vec{u}$ and $U_x = u_x$ that $U_y \le u_y$,
\begin{equation} \label{conf-score}
 A (  \vec{u} , u) : = \Pr( U_y \le u_y | U_x = u_x, \vec{u} ).
\end{equation}

To apply Algorithm \ref{dfn56}, $a_{m+1} (y) $ is conformity score (\ref{conf-score}) for $u_{m+1} = (u_x, u_y)$ and training set $\{ u_1 \cdots u_m \}$,
which we numerically evaluate with  mixture density (\ref{mix3}).
However for $i = 1 \cdots m$,
 $a_{i} (y) $ is conformity score (\ref{conf-score}) for $u_{m+1} = u_i$ and training set
$ \vec{u}^{(i)} \cup \{  u_{m+1}  \}$, with $\vec{u}^{(i)}  = \{ u_1, \cdots,  u_{i-1}, u_{i+1}, \cdots, u_m \}$.
We evaluate $a_{i} (y)$ by generating a mixture density (\ref{mix3}) 
that  approximates posterior predictive density $f(u  | \vec{u}^{(i)} \cup \{  u_{m+1} \}  )$.

\medskip
Note that for $m \rightarrow \infty$ the posterior predictive distribution $f(u | \vec{u})$ identifies with $\tilde{\pi} (u)$,
in which case the conformity score in (\ref{conf-score}) for $u \sim \tilde{\pi} (u)$ is a $U[0,1]$ random variable
and thus the conformity score identifies with its significance level in Algorithm \ref{dfn56},  $a_{m+1} (y) = p(y)$. 
Thereby implying that the bottom end of the $1 - \alpha$ Conformal Prediction Set for $y_{m+1}$ is the $\alpha$ quantile
of the conditional posterior predictive distribution of $U_y | U_x = u_x$.

In the top row of Figure 7 we present  $U[0,1]$ qqplots for the sequence of conformity scores, between each of the $m$ observations and the
training set consisting of the remaining $m-1$ observation, for the simulated examples shown in figure 6. 
For $m = 1000$, we see that  the distribution of the conformity scores for the observations is very close to $U[0,1]$.
Thus the ends of the $0.90$ Conformal Prediction sets would be very close to the dashed orange curves in the bottom plot in Figure 6.

In the bottom of figure 7, we display the Conformal Prediction sets for the $m = 100$ simulated example in Figure 6.
The lower purple dashed curve marks the bottom ends of the $0.95$ Conformal Prediction Sets for $y_{m+1}$ for all $u_x \in [0,1]$,
for conformity score $\Pr( U_y \le u_y | U_x = u_x, \vec{u} )$.
The upper purple dashed curve marks the top ends of the $0.95$ Conformal Prediction Sets for $y_{m+1}$ for all $u_x \in [0,1]$,
for conformity score $\Pr( U_y \ge u_y | U_x = u_x, \vec{u} )$.
Thus the region between the two purple curves forms a  $0.95$ Conformal Prediction Set for $u_{m+1}$.

\begin{figure}[h] 
\centering
\includegraphics[width=1\textwidth,height=.5\textwidth]{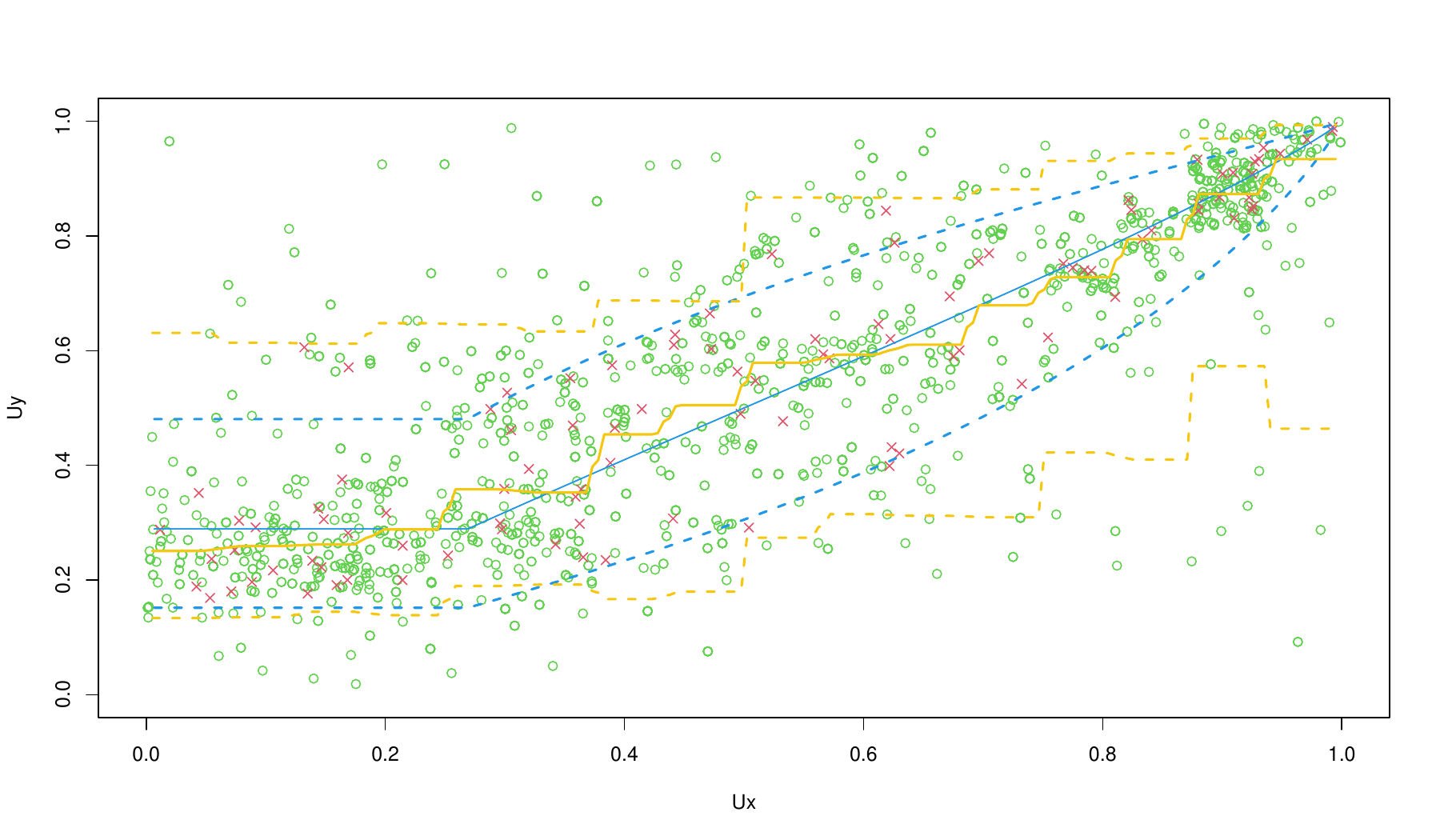}
\includegraphics[width=1\textwidth,height=.5\textwidth]{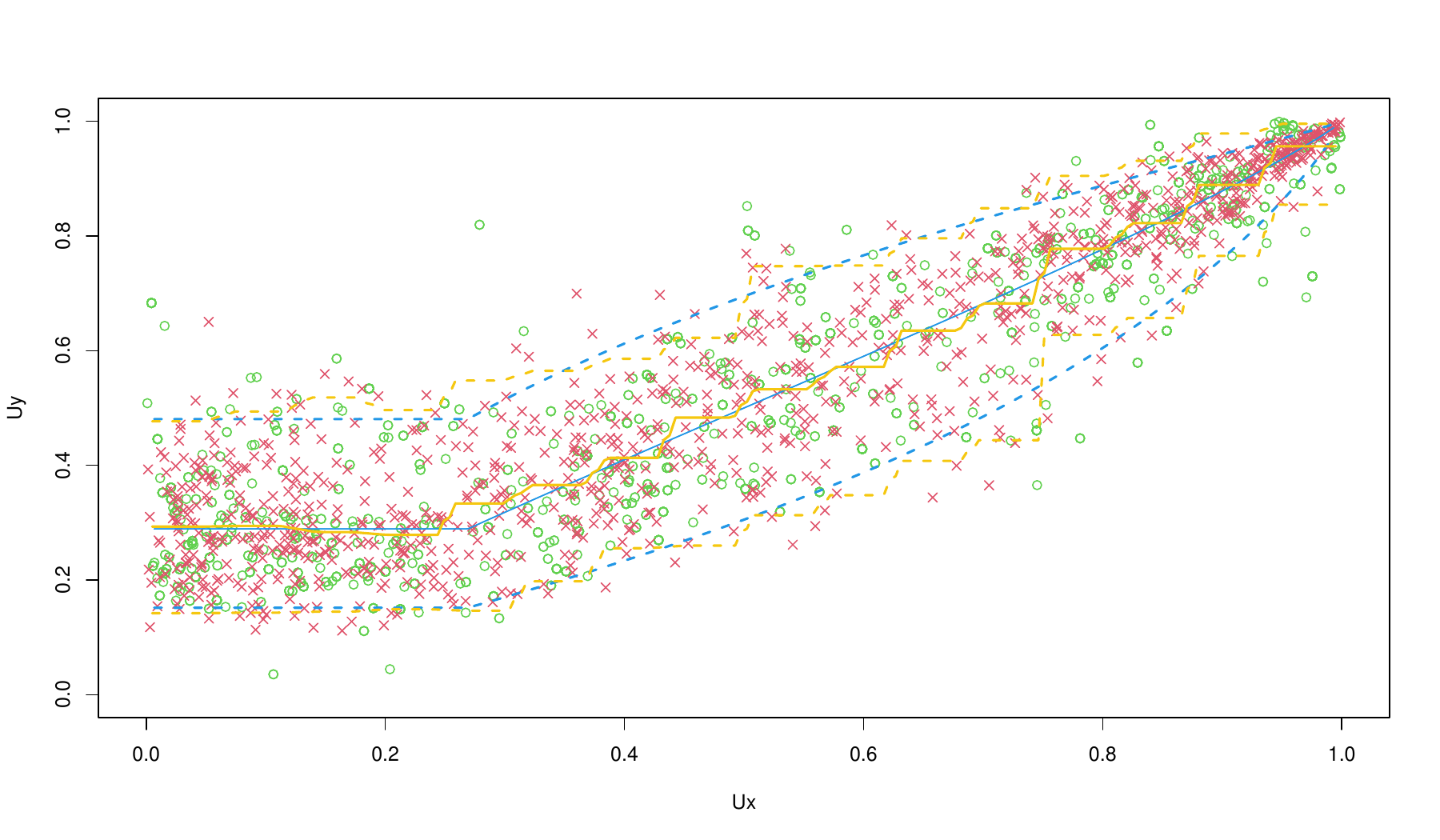}
\caption{Quantile regression simulated example -- Credible Prediction Sets.
The plots correspond to simulations with sample size $m=100$ (top) and $m = 1000$ (bottom).
The Red X's mark the observations, $u_1 \cdots u_m$.
The Green circles are posterior predictive samples, $u^1_{m+1} \cdots u^{2000}_{m+1}$.
The solid Blue curves mark the conditional median of $U_y$ given $U_x$,
the dashed blue curves mark the $0.05$ and $0.95$ quantiles of the conditional distribution of $U_y$ given $U_x$.
The Orange curves are the the $0.05$ and $0.95$ quantiles and median of the conditional posterior predictive distribution of $U_y$ given $U_x$.
}
\end{figure}

\begin{figure}[h] 
\centering
\includegraphics[width=1\textwidth,height=.45\textwidth]{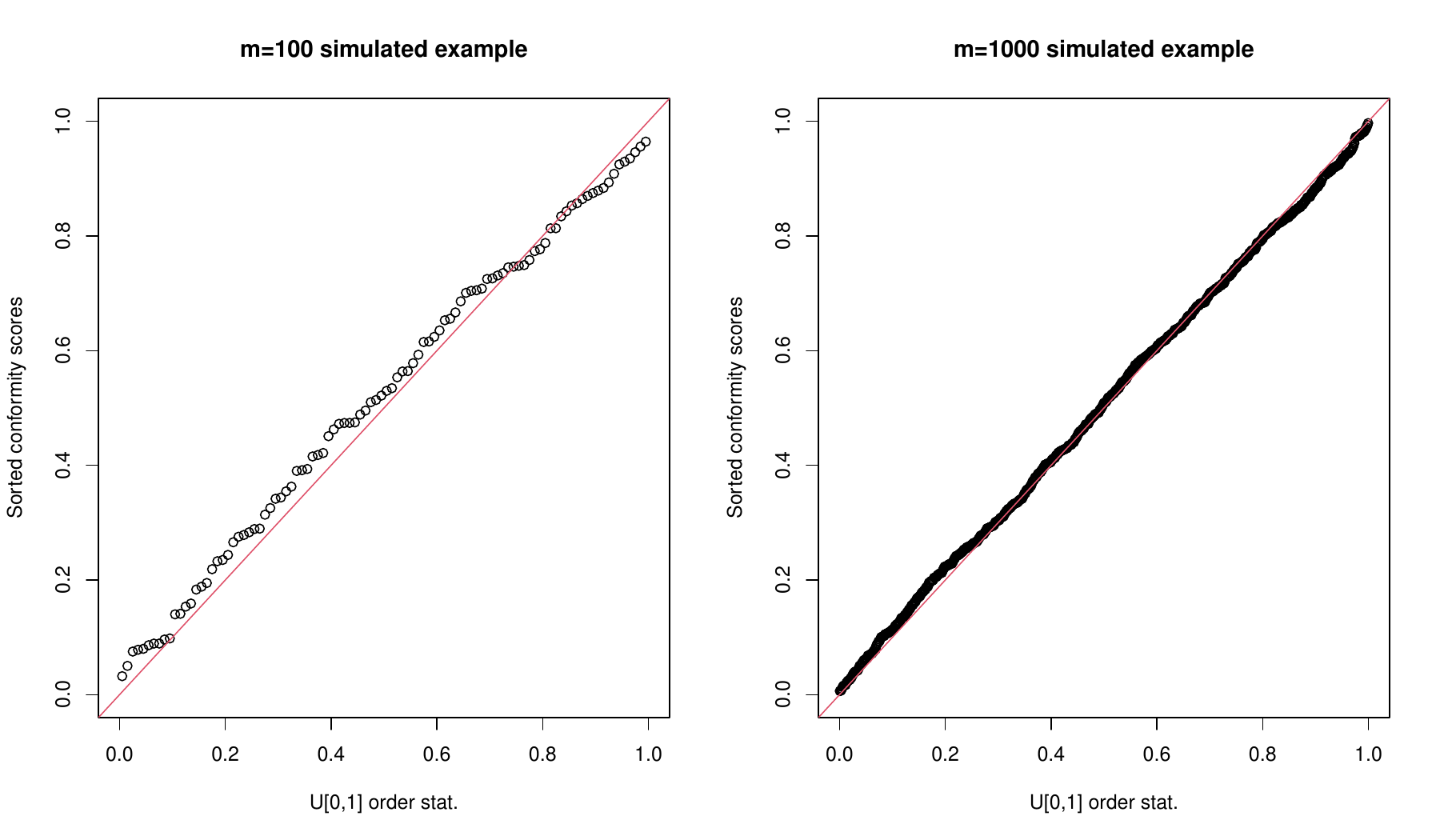}
\includegraphics[width=1\textwidth,height=.55\textwidth]{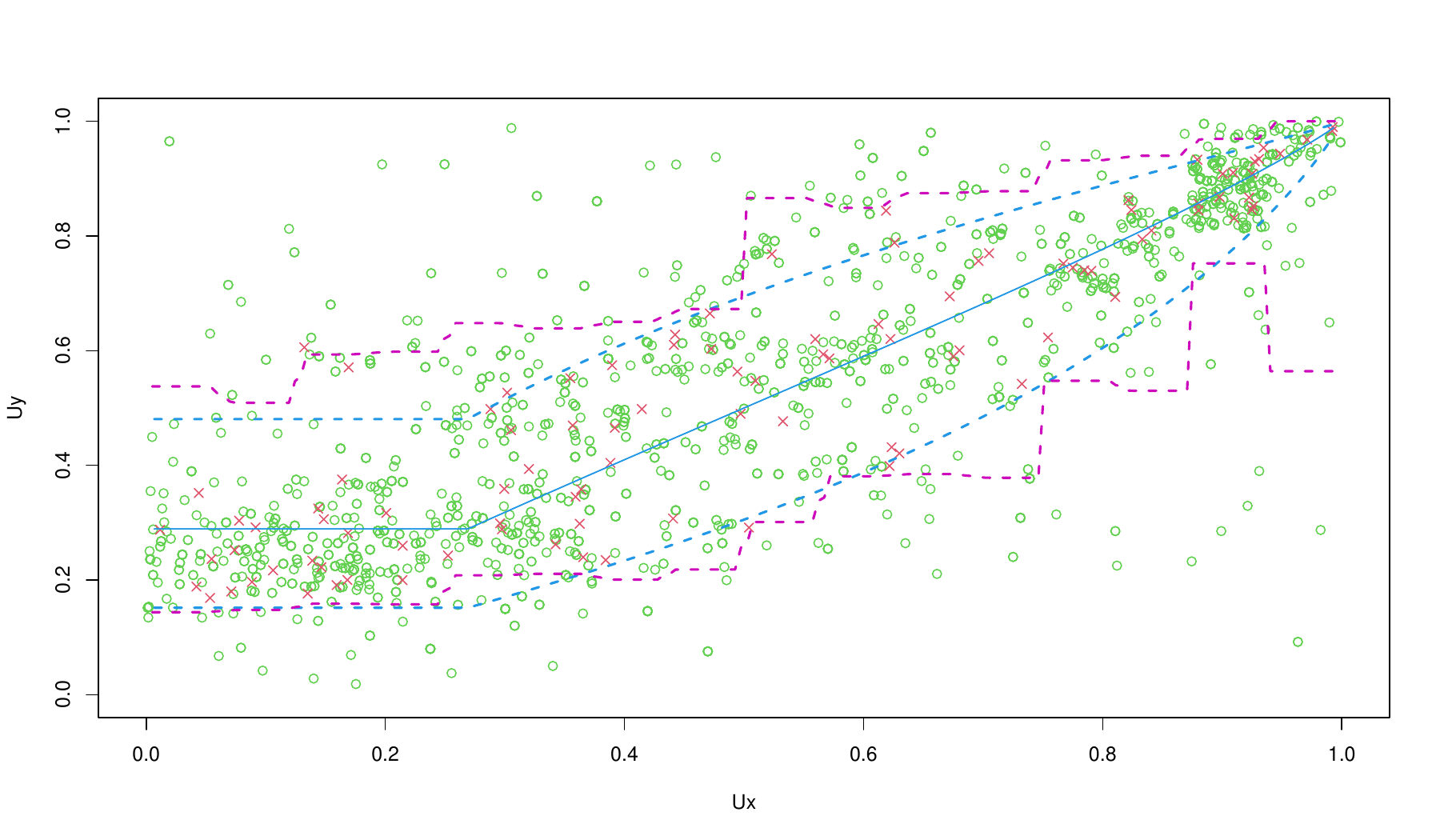}
\caption{Quantile regression simulated example -- Conformal Prediction Sets.
The plots in  the top row are  $U[0,1]$ qqplots for the conformity scores of the $m$ observations 
for the simulated examples shown in Figure 6. 
The Purple dashed curves in the bottom plot are  $0.95$ Conformal Prediction Set for $u_{m+1}$ for the simulated example shown 
in the top plot of Figure 6 (the Red and Green points and blue curves are the same as in Figure 6).
}
\end{figure}

 \bigskip
 \section{High Dimensional Predictive Inference simulation}
 
 In this section, $\tilde{\pi}$ is the joint distribution of 
a categorical variable  $X$,
 which takes on values ``a'', ``b'', ``c'' with probabilities $0.5$, $0.3$, $0.2$, 
 and a continuous $8$-dimensional  random vector $\vec{Y} = ( Y_1 \cdots Y_8)$,
with conditional distribution  $\vec{Y} | X = x \sim N( \vec{\mu}_{x}, \Sigma_{x})$.
 $\vec{\mu}_{x} = ( 0, \cdots, 0)'$ and or all values of $X$;
for $X  = b$ or $X =  c$, $\Sigma_{x} = I_{8 \times 8}$;
 for $X  = a$, $cov(Y_1, Y_2) = 0.8$, 
 other than that  $\Sigma_{x} = I_{8 \times 8}$.
 
 \medskip
In the simulated example, we generate $m =400$ iid  realizations of $\tilde{\pi} ( x, \vec{y})$, 
 $(x_i, \vec{y}_i) = (x_i, y_{i, 1}, \cdots, y_{i, 8})$  for $i = 1 \cdots m$.
 Our goal is to produce $n = 1000$ posterior predictive samples,  
 $(x^1_{m+1}, \vec{y}^1_{m+1}),\cdots,$ $   (x^n_{m+1}, \vec{y}^n_{m+1})$.

To apply our methodology,
for $i = 1 \cdots m$
$(x_i, \vec{y}_i)$ is transformed to $\vec{u}_i = ( u_{i, 1} \cdots u_{i, 10})$ as follows.
 For $j = 1 \cdots 8$, 
 the support of continuous variable $Y_j$ is inflated by $2 \%$ and partitioned into $16$ subintervals.
For $l = 0 \cdots 16$, let $q_{l, j}$ denote the $l / 16$ quantile of $( y_{1, j}, \cdots, y_{m, j})$.
 The marginal subintervals are then  defined:
 $I^j_1  = [ q_{0, j} - (q_{16, j} - q_{0, j}) / 100, q_{1, j}]$,
 $I^j_l  = [ q_{l-1, j}, q_{l, j}]$ for $l = 2 \cdots 15$,
 $I^j_{16}  = [ q_{15, j}, q_{16, j}  +  (q_{16, j} - q_{0, j}) / 100 ]$.
 For $i = 1 \cdots m$, we set $u_{i, j}  = 1/32 + (l-1) / 16$ for $y_{i, j } \in I^j_l$.
 Thus $u_{1,j} \cdots u_{m,j}$ take on the values  $\{ \frac{1}{32}, \frac{3}{32}, \cdots, \frac{31}{32} \}$ with probabilities $1/16$.
The categorical variable $x_i$ is encoded by two dichotomous variables:
$u_{i, 9} = 1/4$ for $x_{i} \ne b$ and  $u_{i, 9} = 3/4$ for $x_{i, 9}  = b$;
$u_{i, 10} = 1/4$ for $x_{i} \ne c$ and  $u_{i, 10} = 3/4$ for $x_{i, 9}  = c$.

The sample $\vec{u}_1 \cdots \vec{u}_m$ is segmented with $L = 10$ dyadic segmentations of $[0,1]^{10}$,
where for $j  = 1 \cdots 8$ variable $U_j$ may be segmented $4$ times in halve 
and for $j  = 9, 10$ variable $U_j$ may be segmented in halve once.
${\cal D}$ is the set of  $1960 = { 8 \choose 4} \cdot { 8 \choose 2}$ segmentations, in which $[0,1]^{10}$ is first partitioned according to $U_{10}$ and $U_{9}$
and then  according to either $U_{j_1}$ or  $U_{ j_2}$ -- four partitions in each dimension --  for all   $j_1  \ne  j_2$ in $\{ 1 \cdots 8 \}$.
For the hBeta model we set $a_0 = 1$.

For each segmentation $\vec{d}$, 
our method produces  posterior predictive samples for $4$ variables, $(u_{m+1, 10},$ 
$u_{m+1, 9}, u_{m+1, j_1}, u_{m+1, j_2})$.
As our model explicitly assumes a Uniform density within each segmentation, 
the values of the remaining $6$ continuous variables, $u_{ m+1, j}$ for $j = 1 \cdots 8$ with $j \ne j_1$ and $j \ne j_2$, 
is sampled with equal probabilities from $\{ \frac{1}{32}, \frac{3}{32}, \cdots, \frac{31}{32} \}$ .
The posterior predictive sample, $( u^1_{m+1, 1} \cdots  u^1_{m+1, 10}), \cdots, (u^n_{m+1, 1} \cdots  u^n_{m+1, 10})$,
is then generated by importance sampling on $\vec{d} \in {\cal D}$.
Lastly, we produce the posterior predictive sample, $( x^1_{m+1}, \vec{y}^1), \cdots, (x^n_{m+1}, \vec{y}^n)$ as follows.
$x^i_{m+1}  = b$ if $u^i_{m+1, 9}  = 3/4$, $x^i_{m+1} = b$ if $u^i_{m+1, 10}  = 3/4$, otherwise $x^i_{m+1} = a$.
While, for $j = 1 \cdots 8$ and $i = 1 \cdots n$, $y^{i}_{m+1, j}$ is sampled from the Uniform distribution on subinterval 
$I^j_k$, with $k =  (u^i_{m+1, j} \cdot 32 +1)/2$.

Note that the normalizing transformation applied to the continuous variables, 
ensures that the marginal posterior predictive distribution of $Y_j$, $j = 1 \cdots 8$,
 will be similar to the marginal distribution of  $Y_j$ even if $U_j$  is not included in $\vec{d}$. 
However, to ensure that the marginal posterior predictive distribution of $X$ will be similar to the marginal distribution of 
$X$, variables $U_{10}$ and $U_9$ had to be included in each segmentation.

\medskip
In Figure 8 we display the logarithm of numerator of Expression (\ref{eq671}), which is proportional to the segmentation probability $\Pr ( \vec{d}   | \vec{u})$,
for the $1960$ segmentations in ${\cal D}$.
The plot reveals that only the segmentations with continuous variables $Y_1$ and $Y_2$ had non-negligible selection probabilities.
Thereby implying that  the posterior predictive samples of $Y_3 \cdots Y_8$ were generated independently from a marginal distribution 
that is very similar to that of the original sample.

For all $\vec{d} \in {\cal D}$ we set $d_1 = 10$ and $d_2 = 9$,
because according to our intuition beginning the segmentation with $U_{10}$ and $U_9$ is supposed to yield the largest segmentation probability.
To test our intuition, $\forall \vec{d} \in {\cal D}$, we compare the segmentation probability of $\vec{d}  = (d_1 \cdots d_{10})$
to the segmentation probability of $\vec{d}' = (d'_1 \cdots d'_{10})$ 
derived by setting $d'_j = d_{j+2}$ for $j = 1 \cdots 8$,   $d'_9 = 10$, $d'_{10} = 9$.
And indeed, for all $\vec{d}$ this change decreased the logarithm of the segmentation probabilities by more than $100$.

The sample of observations, $x_{1} \cdots x_{400}$, 
consisted of $216$ observations with $X = a$, $111$ observations with $X = b$,  $73$ observations with $X = c$.
The posterior predictive sample of the categorical variable, $x^1_{401} \cdots x^{1000}_{401}$, consisted of $535$ observations with $X = a$,
 $327$ observations with $X = b$,  $138$ observations with $X = c$.
In Figure 9 we compare the joint conditional distribution of $Y_1$ and $Y_2$, given $X = a$
and $X \ne a$, for the original sample and for the posterior predictive sample.
Figure 9 reveals that our method managed to pick up the relation between $X$, $Y_1$ and $Y_2$, in $\tilde{\pi}$.
While according to the results shown in Figure 8,
our method also correctly modelled the independence in $\tilde{\pi}$ between $Y_j$, $j = 3 \cdots 8$, and the other variables.

\begin{figure}[h!] 
\centering
\includegraphics[width=1\textwidth,height=.57\textwidth]{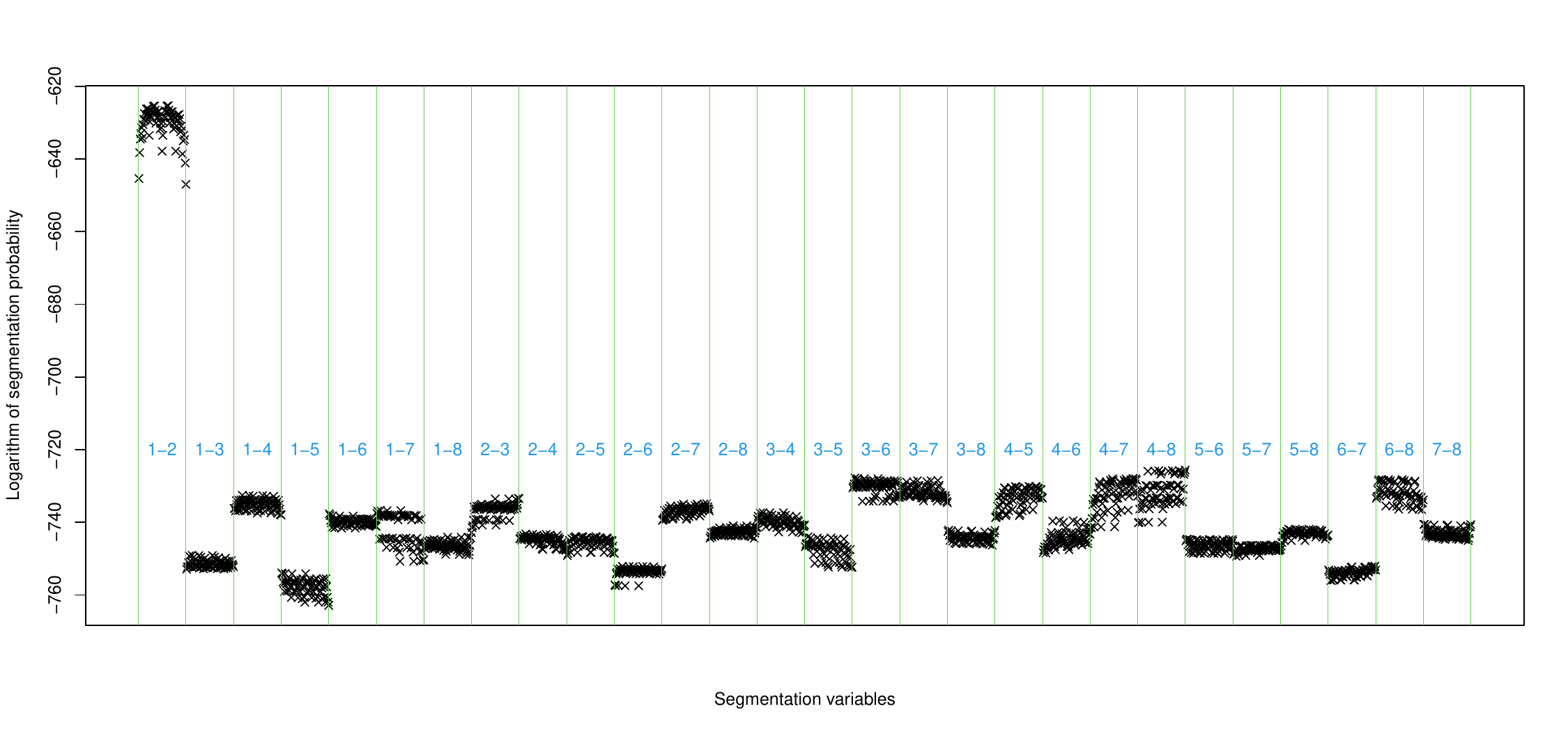}
\caption{High Dimensional Predictive Inference simulation -- Log-probabiliy of the $1960$ segmentations.
The probabilities are arranged in $28$ groups of $70$ segmentations that 
correspond to the pair of continuous variables listed in Blue.
}
\end{figure}

\begin{figure}[h] 
\centering
\includegraphics[width=1\textwidth,height=.5\textwidth]{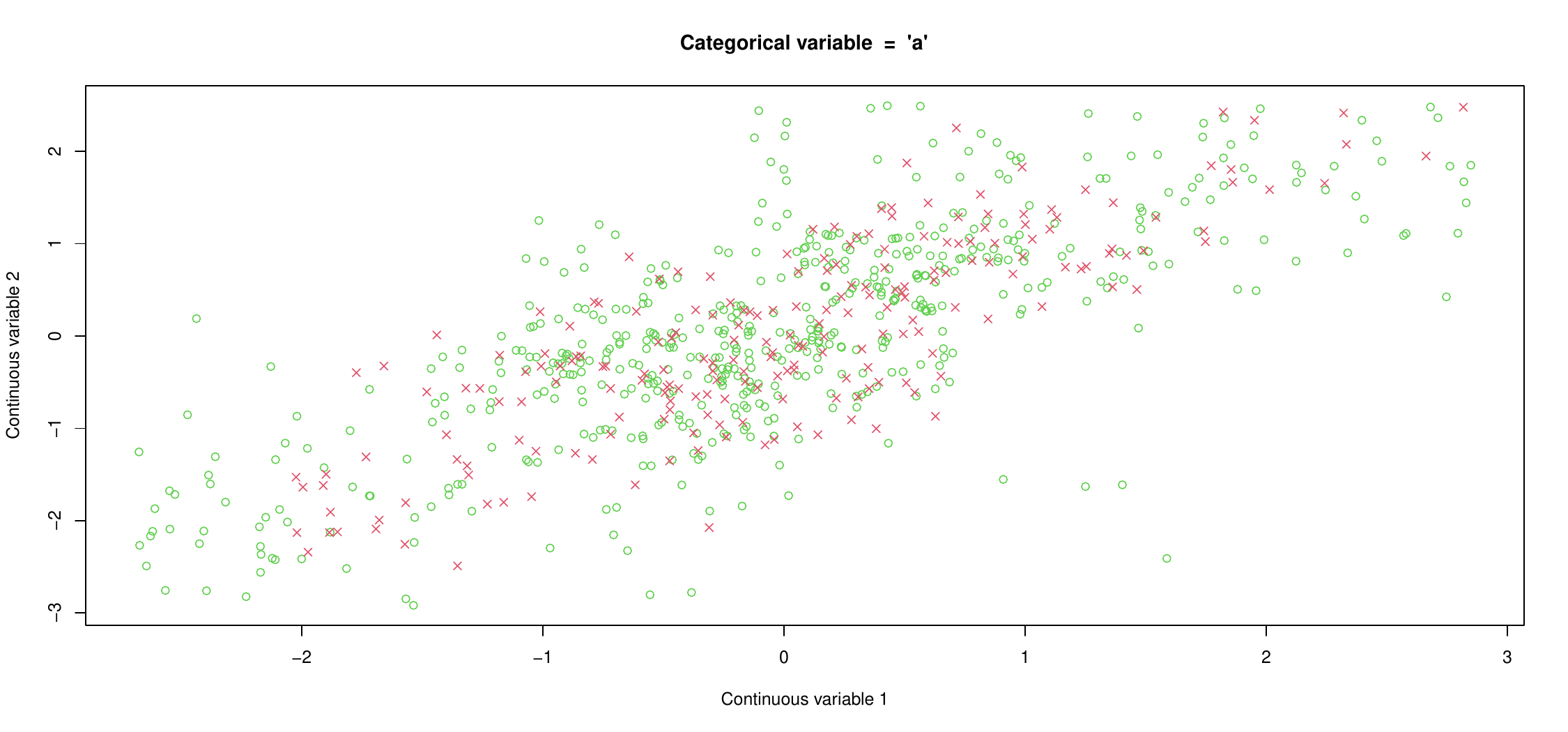}
\includegraphics[width=1\textwidth,height=.5\textwidth]{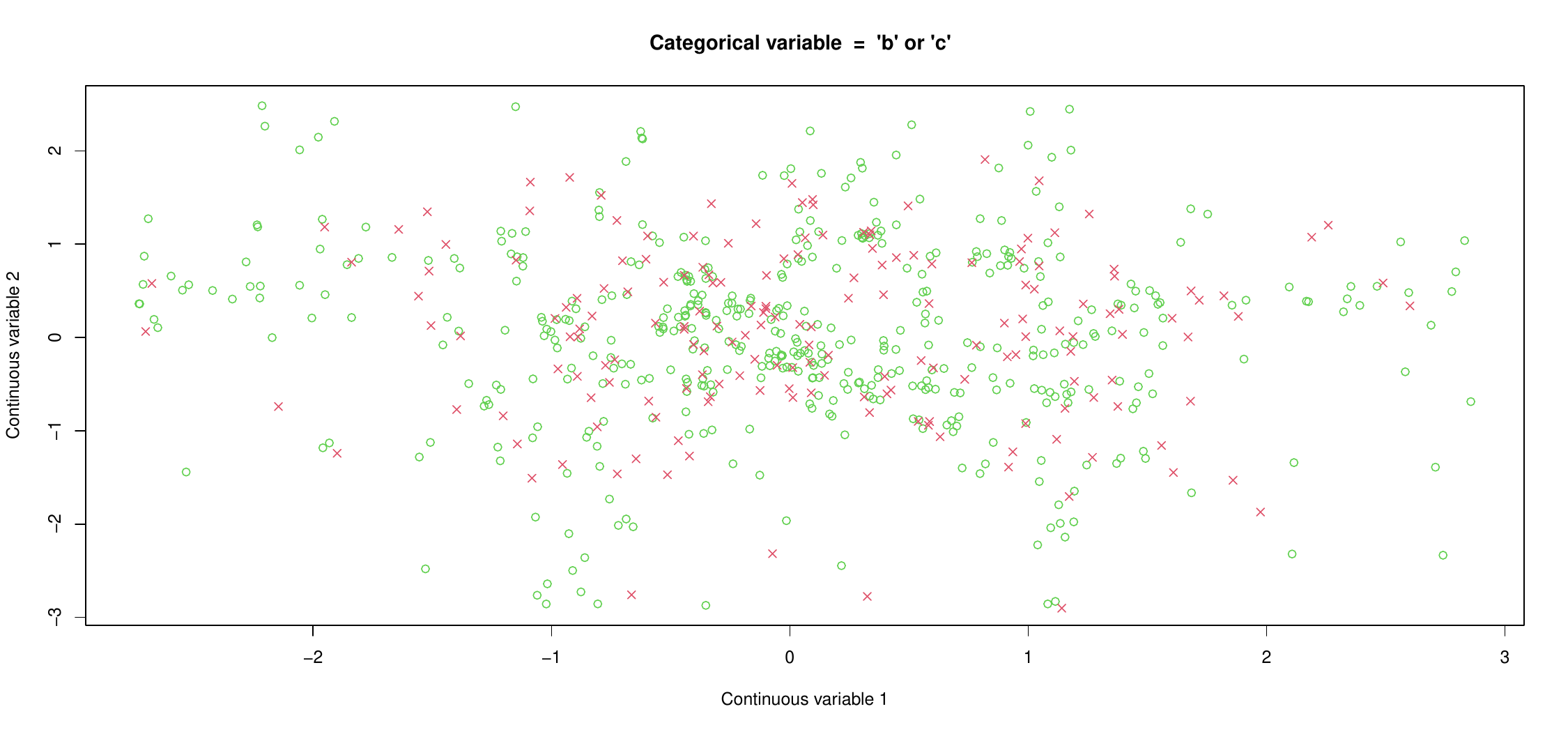}
\caption{High Dimensional Predictive Inference simulation -- joint distribution of $X$, $Y_1$ and $Y_2$.
The Red x signs form scatterplots for the first and second components of  $\vec{y}_1 \cdots \vec{y}_{400}$,
for $x_i = a$ (top plot),  and for $x_i \ne a$ (bottom plot),  $i = 1 \cdots 400$.
The Green circles  form scatterplots for the first and second components of  the posterior predictive samples, $\vec{y}^1_{401}  \cdots \vec{y}^{1000}_{401}$,
for $x^i_{401} = a$, and for $x^i_{401} \ne a$, $i = 1 \cdots 1000$.
}
\end{figure}



 \bigskip
 \section{Unequal dyadic partitioning}
 
In our hierarchical Bayes framework we modeled distribution  $\tilde{\pi} ({u})$ on ${u} \in [0,1]^P$ 
as a mixture of distributions with step-function PDF, 
\begin{equation} \label{def-step}
f ( u  |  \vec{d}, \vec{\pi}_L )  =  \pi_{L,1} \cdot   I_{L,1} (u ; \vec{d})  / (1/2)^L +  \cdots +
\pi_{L, 2^L} \cdot I_{L, 2^L} (u  ; \vec{d})  / (1/2)^L,
 \end{equation}
 for multiple $L$-level dyadic segmentations specified by the segmentation ordering vector 
 $\vec{d} = \{ d_{l,j})_{l = 1 \cdots L}^{j = 1 \cdots 2^l}$ with $d_{l, j} \in \{ 1, \cdots, P \}$.
 Segmentations consist of $2^{L+1} -1$ subintervals: 
 ${\cal I}_0 = [0,1]^P$, and for $l = 1, \cdots, L$ and $j = 1, \cdots, 2^{l-1}$ , ${\cal I}_{l-1, j}$ is partitioned in halve according to dimension $d_{l,j}$ to form
  ${\cal I}_{l, 2 \cdot j -1}$ and  ${\cal I}_{l, 2 \cdot j}$.
  Thus, by construction, the volume for each level-$l$ subinterval is $\vol ({\cal I}_{l,j}) = 1/2^l$.
We will refer to segmentations in which each subinterval is partitioned into two equal volume 
subintervals  as equal volume segmentations.
  
The probabilities $\vec{\pi}_L  = (\pi_{L, 1}, \cdots, \pi_{L, 2^L})$ are generated by a FPT model with  independent Beta variates, $\phi_{l,j} \sim Beta(a_0,a_0)$,
that correspond to conditional subinterval probabilities,
 $\Pr({\cal I}_{l, 2 \cdot j - 1} | {\cal I}_{l-1, j}) = \phi_{l, j}$ and $\Pr({\cal I}_{l, 2 \cdot j} | {\cal I}_{l-1, j}) = 1 - \phi_{l, j}$. 
Thus prior expectation for each conditional subinterval probability is $0.5 = a_0 / (a_0 + a_0)$,
and the prior expectation of $\pi_{L,j}$  is $1/2^L$.
Dividing indicator functions $I_{L,j}$  in (\ref{def-step}) by $\vol ({\cal I}_{L,j}) = 1/2^L$,
implies that the prior expectation of $f ( u  |  \vec{d}, \vec{\pi}_L )$
is the uniform PDF on $[0,1]^P$.

\medskip
To implement our modeling approach to random vector, $\vec{X} = (X_1, \cdots, X_P)$, $X_p \in \mathbb{R}$.
For $p = 1, \cdots, P$, we specify mapping $\omega_p$ that maps   $X_p$ into $[0,1]$.
In principle, we want the marginal distribution of $U_p = \omega_p (X_p)$ to be $U[0,1]$.
The first reason for this is that our method shrinks the density estimates to the uniform prior distribution of the generative model.
Thus if the marginal distribution of the variates is non-uniform then even if the variable are independent our shrinkage estimates are invariably biased.
The second reason is that we conjecture that our model works well because it assign large posterior probabilities to  segmentations that induce a monotone ordering  
for $\tilde{\pi} (u)$,
which may be well approximated by step-functions and is easy to estimate by FPT modeling. 
In principle, having segmentations that provide a good approximation and induce a monotone ordering for $\tilde{\pi} (u)$ 
is possible if we consider a huge number of segmentations with very large $L$.
However approximating a density with equal volume dyadic segmentations with relatively small $L$ is more effective if the marginal distribution of its components is uniform.
This requirement is formalized in Zhang (2019) that shows that dyadic partitioning provides uniformly consistent modeling for the bivariate copula.
In this note we address the problem of how to implement our modeling approach with unequal volume dyadic partitions,
in cases where it makes more sense in the application to partition continuous $X_p$ into unequally probable subintervals and in cases where $X_p$ is not a continuous variable.


\bigskip
\subsection{Unequal volume dyadic partitions for continuous variables}
To implement our modeling approach to $m$ samples of continuous random vector, $\vec{X} = (X_1, \cdots, X_P)$,
$\vec{x}_1, \cdots, \vec{x}_m$  with $\vec{x}_k = (x_{k1}, \cdots, x_{kP})$.
We specify minimal value, $x_{0, p}$ and  maximal value, $x_{\infty, p}$, for $X_p$,
and define a continuous empirical CDF mapping,
\begin{equation} \label{def-map}
\omega_p (x) =  \begin{cases*}
                    \frac{1}{m+1} \cdot \frac{ x - x_{0, p}}{ x_{(1), p} -  x_{0, p}}   & if  $x \in [ x_{0, p}, x_{(1), p}]$  \\
                    \frac{1}{m+1} \cdot \left( k +  \frac{ x - x_{(k), p}}{ x_{(k+1), p} -  x_{(k), p}} \right)   & if  $x \in ( x_{(k), p}, x_{(k+1), p}]$ for $k = 1, \cdots, m-1$  \\
                    \frac{1}{m+1} \cdot \left( m  +  \frac{ x - x_{(m), p}}{ x_{\infty, p} -  x_{(m), p}} \right)   & if  $x \in ( x_{(m), p}, x_{\infty, p}]$ \\
                 \end{cases*} 
\end{equation}
where $x_{(1), p} < \cdots < x_{(m), p}$ are the order statistics for the sample of $X_p$,  $x_{1 p},  \cdots,  x_{m p}$.
For $p = 1, \cdots, P$ and $k = 1, \cdots, m$,  let  $u_{k p} = \omega_p (x_{k p})$.
Given $u_1, \cdots,  u_{m}$ with  $u_k = (u_{k1}, \cdots, u_{kP})$,
we may use our method to generate posterior predictive samples $u^{1}_{m+1}, \cdots, u^n_{m+1}$.
For $i = 1, \cdots, n$, 
applying the inverse mapping,  $\omega^{-1}_p: [0,1] \rightarrow \mathbb{R}$,
to the components of $u^i_{m+1} = (u^i_{m+1, 1}, \cdots, u^i_{m+1, P})$,
yields posterior predictive samples $\vec{x}^{1}_{m+1}, \cdots, \vec{x}^{n}_{m+1}$. 

\medskip
In this subsection we further assume that we are given $L_p$, the number of times variable $X_p$ can be partitioned,
and the points $\{ x^p_{l,j} \}_{l = 1, \cdots, L_p}^{j = 1, \cdots, 2^l-1}$ at which the support of $X_p$, $[x_{0, p}, x_{\infty, p}]$, can  be partitioned.
We use $u^p_{l,j} = \omega_p (x^p_{l,h})$ for  $l = 1, \cdots, L_p$ and $j = 1, \cdots, 2^l-1$, $u^p_{l,0} = 0$ and $u^p_{l,2^l} = 1$, 
for specifying the unequal partitioning of $U_p$.
Let ${\cal I}^p_{0,1} = [0,1]$ and define the level  $l = 1, \cdots, L_p$,
marginal unequal segmentation of $U_p$:
\[  
{\cal I}^p_{l,1} = [u^p_{l,0}, u^p_{l,1}] , \;  \;  {\cal I}^p_{l,2} =  ( u^p_{l,1}, u^p_{l,2}] \ , \; \cdots, \;
\;   {\cal I}^p_{l,2^l-1} =  ( u^p_{l,2^l-2}, u^p_{l,2^l-1}] \ , \; \;   {\cal I}^p_{l,2^l} =  ( u^p_{l,2^l-1}, u^p_{l,2^l}].
\]
We use this marginal segmentation to construct the dyadic segmentation $\{ {\cal I}_{l,j} (\vec{d} )\}_{l = 1, \cdots, L}^{j = 1, \cdots, 2^l}$.
If $d_{l,j} = p$ and the range of $U_p$ in ${\cal I}_{l-1, j}$ is ${\cal I}^p_{l_p-1,j_p}$ then we partition ${\cal I}_{l-1, j}$ into
\[
{\cal I}_{l, 2 \cdot j -1} =  {\cal I}_{l-1, j} \cap \{ u : \; u_p \in {\cal I}^p_{l_p,2 \cdot j_p -1} \}   \; \; \; \hbox{and} \; \; \;   
{\cal I}_{l, 2 \cdot j } =  {\cal I}_{l-1, j} \cap \{ u :  \; u_p \in  {\cal I}^p_{l_p,2 \cdot j_p} \}.
\]
Thus for ${\cal I}_{l, j} = \cap_{p = 1}^P \{ u : u_p \in  {\cal I}^p_{l_p,j_p} \}$,
\[
\vol ( {\cal I}_{l,j} ) = \Pi_{p = 1}^P  (  u^p_{l_p,j_p} - u^p_{l_p,j_p-1} ).
\]
The statistics for our modeling approach are the subinterval counts,
\[
N_{L,j} ( \vec{u}, \vec{d}) =  \sum_{k = 1}^m I_{L,j} (u_k ; \vec{d}).
\]
Where by construction marginal subinterval volumes are given by the corresponding observation counts,
\[
N^p_{l_p,j_p} ( \vec{u}, \vec{d})   := | \{ k:  u_{k p}  \in (  u^p_{l_p,j_p-1},  u^p_{l_p,j_p} ] \} |  =  \lfloor (m+1) \cdot (  u^p_{l_p,j_p} - u^p_{l_p,j_p-1} ) \rfloor.
\]

\medskip
The changes to the dyadic segmentations require us to make changes in the FPT model in the generative model.
At each node we change the Beta prior hyper-parameters in the generative model from fixed values $a_0$,
to adaptive prior hyper-parameter values $\{ a_{l,j}, b_{l,j} \}_{l = 1 \cdots j}^{j = 1 \cdots 2^{l-1}}$ 
that are proportional to the subinterval volumes.
For each $\vec{d}$, $l = 1, \cdots, L$, and $j = 1, \cdots, 2^{l-1}$, the conditional subinterval 
probability $\Pr ( {\cal I}_{l , 2 \cdot j - 1}  ( \vec{d})  |  {\cal I}_{l -1,  j} ( \vec{d}) )$  in the generative model is $\phi_{l, j} \sim Beta ( a_{l,j}, b_{l,j})$ with 
\begin{equation} \label{def-hyper}
a_{l,j}   / ( a_{l,j}  + b_{l,j})  = \vol ({\cal I}_{l,2 \cdot j - 1}) /  \vol ( {\cal I}_{l-1, j }).
 \end{equation}
Noting that expression (\ref{def-hyper}) is also the prior expectation of the conditional subinterval probability, 
$\Pr({\cal I}_{l, 2 \cdot j - 1} | {\cal I}_{l-1, j})  = \phi_{l, 2 \cdot j - 1}$,
then for $j'  = 1, \cdots, 2^L$ the prior expectation of subinterval probability $\Pr({\cal I}_{L,  j'}) = \pi_{L, j'}$ is $\vol ( {\cal I}_{L,  j'})$.
To ensure that the prior expectation of $f ( u  |  \vec{d}, \vec{\pi}_L )$ is the uniform PDF on $[0,1]^P$, 
we define the step function density produced by the generative model,
\begin{equation} \label{def-step-2}
f ( u  |  \vec{d}, \vec{\pi}_L )  =  \pi_{L,1} \cdot   I_{L,1} (u ; \vec{d})  /  \vol ({\cal I}_{L,1})  +  \cdots +
\pi_{L, 2^L} / \vol ({\cal I}_{L,2^L}).
 \end{equation}

\bigskip
\subsection{Unequal volume dyadic partitions for ordinal variables}

In this subsection we assume that variable $X_p$ can take on $m_p$ given ordered values, $v_1 < v_2 < \cdots < v_{m_p}$.
We are also given, $L_p$,  and the values of $X_p$,  $\{ x^p_{l,j} \}_{l = 1, \cdots, L_p}^{j = 1, \cdots, 2^l-1}$ at which  $X_p$  can  be partitioned.
For  $m$ samples of $X_p$,  $x_{1 p}, \cdots, x_{m p}$, we define a discrete empirical CDF mapping,
\begin{equation} \label{def-map-disc}
\omega_p (x) =  | \{  k : x_{k p} \le x \} |  / m.
\end{equation}

From here we use the same pipeline as the continuous $X_p$ case.
We define, $u_{k p} = \omega_p (x_{k p})$ for $k = 1, \cdots, m$.  
Use $u^p_{l,j} = \omega_p (x^p_{l,h})$, $u^p_{l,0} = 0$ and $u^p_{l,2^l} = 1$, 
for specifying the marginal partitioning of $U_p$, ${\cal I}^p_{l,j} = [u^p_{l,j-1}, u^p_{l,j}]$,
which is then used to construct the dyadic segmentation $\{ {\cal I}_{l,j} (\vec{d} )\}_{l = 1, \cdots, L}^{j = 1, \cdots, 2^l}$.
The changes to the FPT model in the generative model are the same as the continuous $X_p$ case.

Lastly, given posterior predictive samples ${u}^{1}_{m+1, p}, \cdots, u^n_{m+1, p}$,
we apply the inverse mapping to yield posterior predictive samples of $X_p$,
 $x^i_{m+1, p} = \omega^{-1}_p ( u^i_{m+1, p})$ for $i = 1, \cdots, n$.
 Note that while the posterior predictive samples of $U_p$ can take on any value in $[0,1]$,
 the use of the empirical CDF mapping in (\ref{def-map-disc})
ensures that the posterior predictive samples of $X_p$ can only take on values $v_1,  \cdots,  v_{m_p}$.

\bigskip
\begin{exa}
We illustrate unequal volume dyadic partitions for $L=3$ dyadic segmentations of $[0,1]^2$
on a simulated example of $500$ samples of continuous variables $\vec{X} = (X_1, X_2)$.
The first variable is $X = X_1$ that can be partitioned twice, $L_1 = 2$, at $x^1_{1,1} = 1$, $x^1_{2,1} = 0.5$, $x^1_{2.3} = 1.5$.
The second variable is $Y = X_2$ that can be partitioned once, $L_1 = 1$, at $x^1_{1,1} = 0$.

Figure 10 is a scatterplot of $(X_1, X_2)$.
Figure 11 displays the mappings, $\omega_1 = \omega_x$, $\omega_2 = \omega_y$, and inverse mappings,
defined in  (\ref{def-map}) for the simulated data.
For this example we empirically set the maximal and minimal values for the support for variables $X_1 = x$ and $X_2 = Y$,
\begin{eqnarray*}
&& x_{0,1} := x_{(1)} - (x_{(2)} - x_{(1)}) = 0.004 \ , \; \;   x_{\infty, 1} := x_{(500)} + (x_{(500)} - x_{(499)}) = 1.997, \\
&& x_{0,2} :=  y_{(1)} - (y_{(2)} - y_{(1)}) = -4.325 \ , \; \;  x_{\infty, 2} := y_{(500)} + (y_{(500)} - y_{(499)}) = 8.926.
\end{eqnarray*}
Recall that by definition, the maximal and minimal variable values (displayed by the red x's in Figure 2)
 are mapped to $U_p = 0$ and $U_p = 1$.
Applying the mappings  to the values at which $X$ and $Y$ may be partitioned yields:
\begin{eqnarray*}
&& u^1_{2,1}  = \omega_1 (0.5) = 0.240 \ , \; \; u^1_{1,1}  = \omega_1 (1.0) = 0.538 \ , \; \; u^1_{2,3}  = \omega_1 (1.5) = 0.752, \\
&& u^2_{1,1}  = \omega_2 (0) = 0.209,
\end{eqnarray*}
which specify marginal subintervals with volumes that are proportional to observation counts.
Thus, the floor of $501 \cdot (  u^1_{2,3} - u^1_{2,1})  = 258.07$ is equal to the number of observations with 
$u_x$ between $u^1_{2,1}$ and $u^1_{2,3}$, $26 + 123 + 26 + 83 = 258$.
While the the floor of $501 \cdot ( 1 -  u^2_{1,1})  = 396.53$ is equal to the number of observations with 
$u_y$ greater than $u^2_{1,1}$, $89 + 123 + 83 + 101 = 396$.
Figure 12 is a scatterplot of $(U_1, U_2)$, with dashed lines specifying the level $3$ subintervals, and observation counts and volumes for each subinterval.
In (\ref{def-hyper}) we suggest that the odds between the Beta hyper-parameters is specified by the volume odds for the partition.
Example 1: for default segmentation partition of ${\cal I}_{1,1}$ into ${\cal I}_{2,1}$ and ${\cal I}_{2,2}$ the hyper-parameter odds is 
\[
\frac{ a_{2,1}}{a_{2,1} + b_{2,1}} = \frac{0.0501 + 0.1900}{0.0501 + 0.1900 + 0.0622 + 0.2362} = 0.4458.
\]
Example 2: for default segmentation partition of ${\cal I}_{2,3}$ into ${\cal I}_{3,5}$ and  ${\cal I}_{3,6}$ the hyper-parameter odds is 
\[
\frac{ a_{2,3}}{a_{2,3} + b_{2,3}} = \frac{0.0452}{0.0452 + 0.1715} = 0.2085.
\]
\end{exa}

\begin{figure} \label{figure1}
\center
\includegraphics[width=1\textwidth, height=1\textwidth]{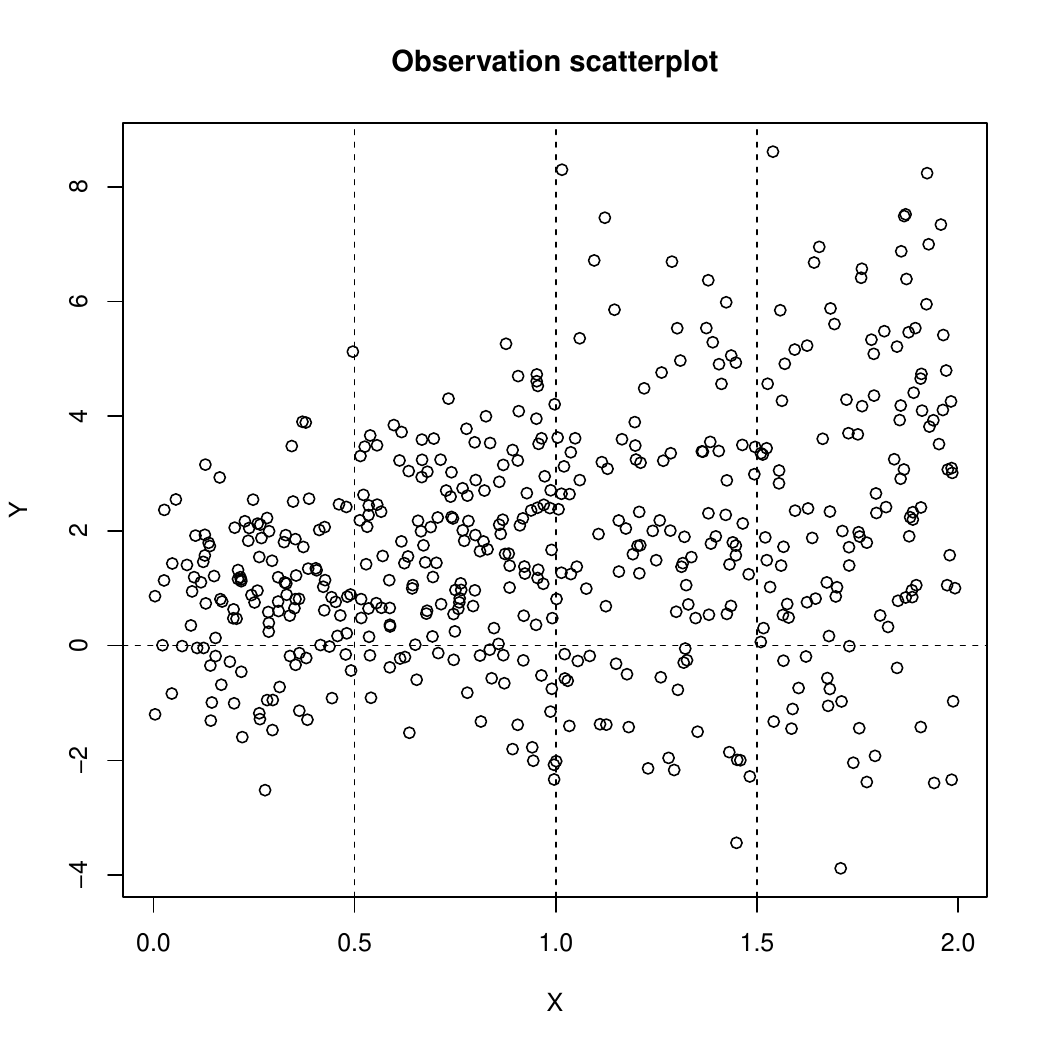}
\caption{$\vec{X} = (X_1, X_2)$ observations scatterplot. The dashed lines mark the points at which the support of the variables is partitioned.}
\end{figure}

\begin{figure} \label{figure2}
\center
\includegraphics[width=1\textwidth, height=1\textwidth]{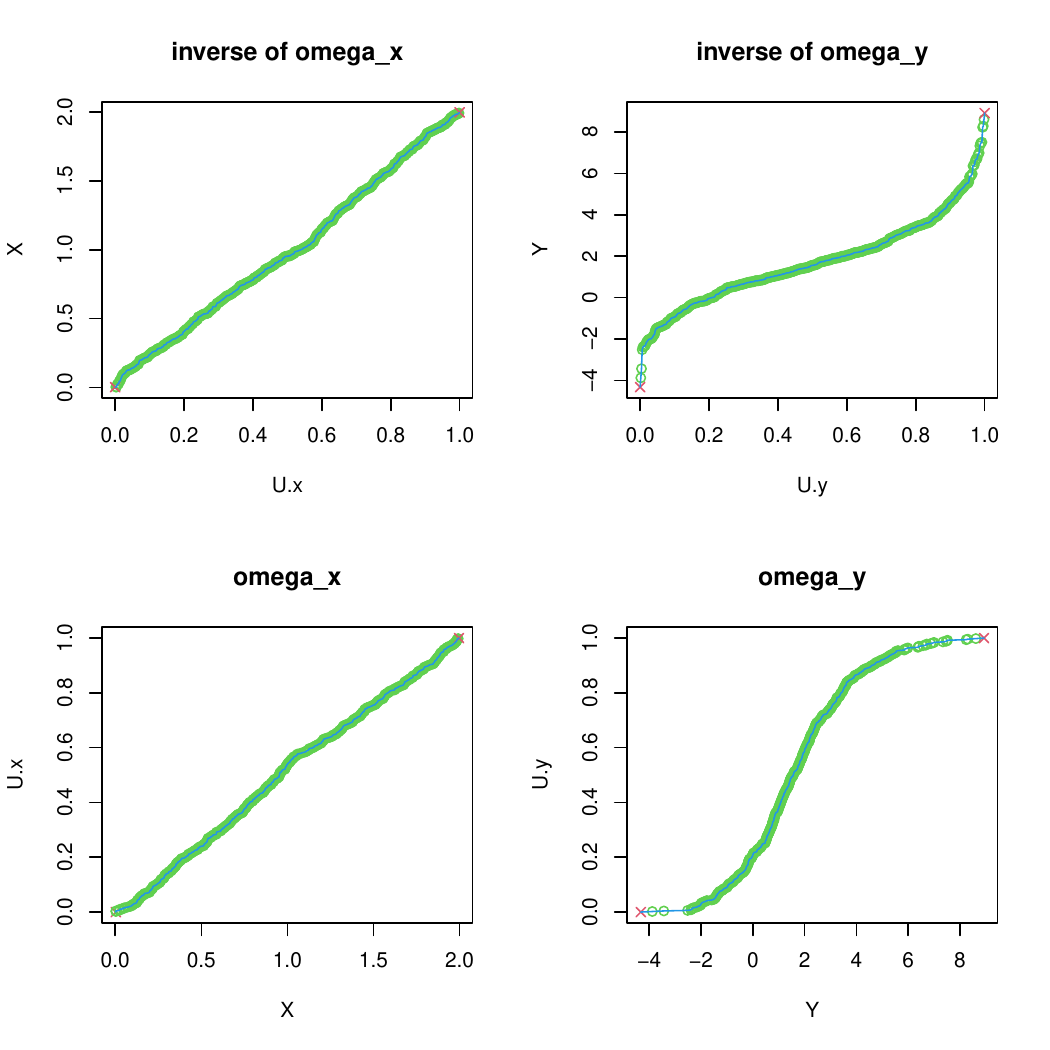}
\caption{Marginal mappings, $\omega_1 (x)$ and $\omega^{-1}_1 (u)$ for $p = 1, 2$, for simulated examples. 
The green circles are $(x_{k p}, u_{k p})$ for $k = 1, \cdots, 500$ and $p = 1,2$.
The red x's are the points corresponding to $x_{0, p}$ and $x_{\infty, p}$ for $p = 1,2$.
The blue curves display the mapping's values $\forall u \in [0,1]$ or $\forall x \in [x_{0,p}, x_{\infty,p}]$.}
\end{figure}

\begin{figure} \label{figure3}
\center
\includegraphics[width=1\textwidth, height=1\textwidth]{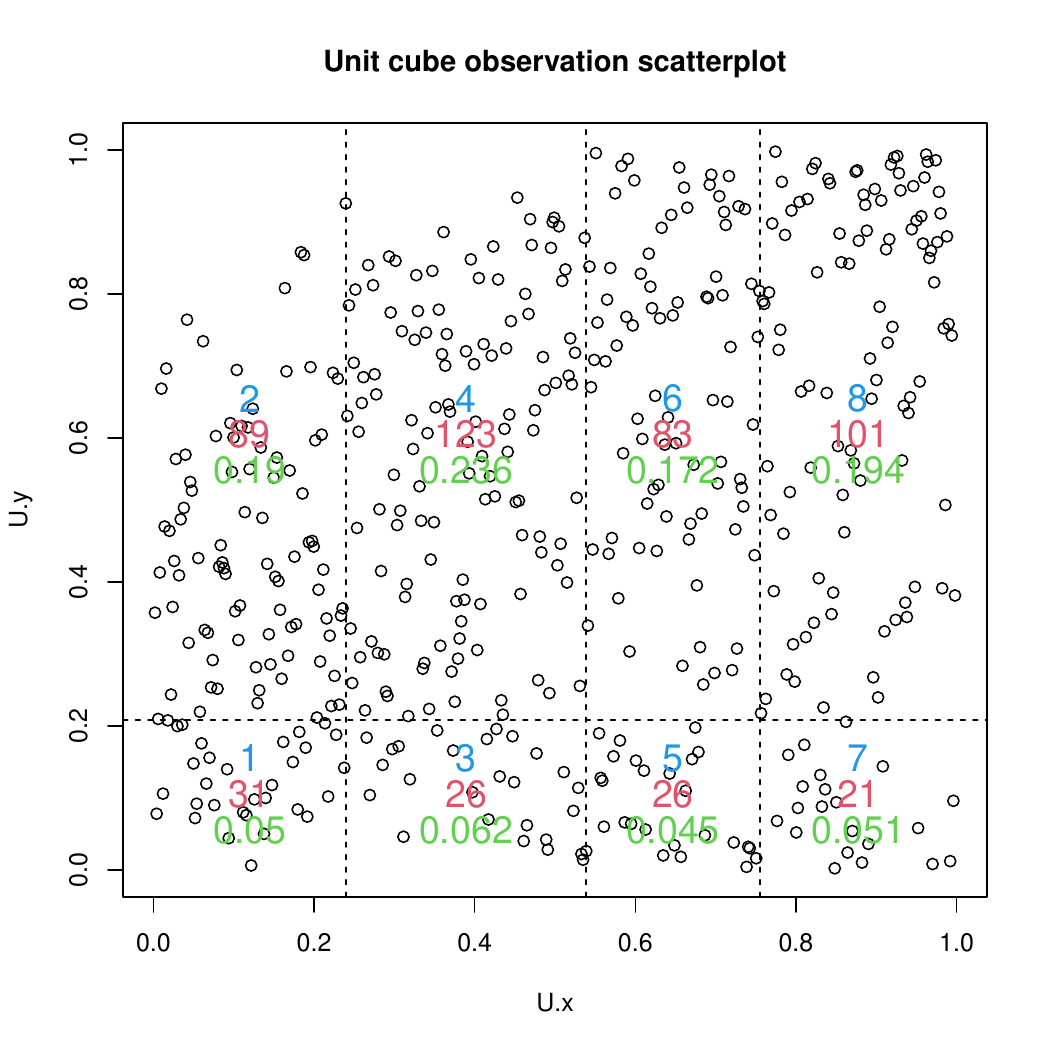}
\caption{${U} = (U_1, U_2)$ observations scatterplot. The dashed lines are drawn at $u^p_{l,j}$.
The blue number is the index $j = 1, \cdots, 8$ of subineterval ${\cal I}_{3,j}$ in the default segmentation $\vec{d} = XXY$.
The red numbers are the subinterval observation counts $N_{3,j}$
The green numbers are the subinterval volumes, $\vol({\cal I}_{3,j})$.}
\end{figure}

 \bigskip
 \section{Discussion}
 
We presented a Bayesian framework for density estimation and predictive inference that is based on multiple random dyadic segmentations of the P-dimensional unit cube.
To apply our methodology to a sample of $m$ realizations of random vector $\vec{X}  \in  \mathbb{R}^P$,
it is necessary to specify a mapping for each coordinate of $\vec{X}$ into $[0,1]$,
set the value of $a_0$ and the segmentation level $L$,  and specify the set of segmentations ${\cal D}$.

Our working experience suggests setting $a_0 = 1$.
Our focus in this work is recovering the dependence between the components of $\vec{X} = (X_1 \cdots X_P)$. 
For small $P$, a linear mapping between an interval, slightly larger than the support of $X_j$, and $[0,1]$ should be satisfactory.
For larger $P$, the mapping we applied in Section 5, based on the empirical distribution of the variable, 
allowed us to recover the marginal distribution of a continuous variable $X_j$ without including it in the segmentation.

 In Section 6 we propose modifications for the marginal mappings of the components of $\vec{X}$ into $[0,1]$, and the 
 specification of the FPT model Beta distribution prior hyper parameter values and step-function PDF, for the case 
 that a continuous $X_p$ is partitioned into unequally probable subintervals or  $X_p$ is an ordinal categorical variable.
For equal volume segmentation we use the same value $a_0$ for the the two Beta prior hyper parameters.
 For the unequal volume setting we suggest changing the hyper parameters odds from $0.5$ to 
 the corresponding data adaptive volume odds. However it is not clear which specific hyper parameter values to use. 

For the simulated examples, we considered segmentations in which all subintervals at the same level were partitioned according to the same coordinate
(therefore $| {\cal D}|$ was of order $L^P$)
and we evaluated the posterior predictive distribution by importance sampling over all  $\vec{d} \in {\cal D}$.
This is not computationally feasible for large $P$.
 To overcome this problem we plan to consider a richer family of segmentations ${\cal D}$, in which the subintervals in the same level may be partitioned in different directions,
 and to evaluate  the posterior predictive distribution by MCMC algorithms that perform random walks on ${\cal D}$.
 
The most interesting property of our method,  illustrated in the simulated examples,
 is that it favours segmentations for which the step-function density approximates the distribution of $\vec{X}$ well,
 and yields a step-function density that is easy to estimate by a Polya tree.
 We plan to provide theoretical support for this observation, which we will try to formalize as conjectures.
Note that each segmentation $\vec{d}$ of $[0,1]^P$ into the level $L$ subintervals,  ${\cal I}_{L,1} (\vec{d})  \cdots {\cal I}_{L,2^L} (\vec{d})$,
specifies a mapping from $[0,1]^P$ to $\{ 1, 2, \cdots, 2^L \}$  (see Figure 1 and top row of plots in Figure 5).
We conjecture that $\vec{d}$ for which distribution $\tilde{\pi}$ is increasing with respect to this mapping have larger posterior probabilities;
provide smaller approximation error; yield increasing step-function densities that are easier to estimate.

\end{document}